\documentclass[12pt]{article} 

\usepackage{amsmath,amssymb}

\makeatletter
\@addtoreset{equation}{section}
\makeatother
\def\theequation{\thesection\arabic{equation}}

\newcommand{\bra}{\langle}
\newcommand{\ket}{\rangle}
\newcommand{\no}{\nonumber}
\newcommand{\To}{\Rightarrow}
\def\thefootnote{%
\fnsymbol{footnote}}

\begin{document}
\begin{flushright}
NUP-A-06-2
\end{flushright}
\begin{center}
\Large{\bf A Wave-Packet View of \\
Neutrino Oscillation and Pion Decay} \\
\end{center}
\begin{center}
\large{Chikara Fuji}\footnote{E-mail address: t-fuji@phys.ge.cst.nihon-u.ac.jp}
\end{center}
\begin{center}
{\it Department of General Education, Junior College Funabashi Campus \\
Nihon University, Funabashi 274-8501, Japan} \\
\end{center}
\begin{center}
\large{Yasumasa Matsuura, Toshihiro Shibuya and S.Y.Tsai}\footnote{E-mail address: tsai@phys.cst.nihon-u.ac.jp}
\end{center}
\begin{center}
{\it Institute of Quantum Science and Department of Physics} \\
{\it College of Science and Technology, Nihon University, \\
Tokyo 101-8308, Japan}
\end{center}
\begin{abstract}
Kinematical aspects of pion decay $\pi \rightarrow \mu \nu$ is studied, 
with neutrino mixing taken into account. 
An attempt is made to derive the transition probability for such a 
sequence of processes: a $\pi^+$ produced at $(\vec{x}_{\pi},t_{\pi})$ 
with momentum $\vec{p}_{\pi}$ decays into a $\mu^+$ and a $\nu_{\mu}$ 
somewhere in space-time and then the $\mu^+$ is detected at 
$(\vec{x}_{\mu},t_{\mu})$ with momentum $\vec{p}_{\mu}$ and a 
$\nu_{\alpha}$ (a neutrino with flavor $\alpha = e$, $\mu$, $\cdots$) 
is detected at $(\vec{x}_{\nu},t_{\nu})$ with momentum $\vec{p}_{\nu}$. 
It is shown that
\begin{enumerate}
\item
if all the particles involved are treated as plane-waves, that is, 
if each particle is assumed to possess a strictly fixed momentum, 
the energy-momentum conservation would eliminate the neutrino 
oscillating terms, leaving each mass-eigenstate to contribute 
separately to the transition probability;
\item
if one, taking into account that the momenta relevant may not be 
free from some uncertainty (or dispersion), treats all the particles 
involved as wave-packets, the neutrino oscillating terms 
would appear and would be multiplied by two suppression factors, 
which result from distinction in velocity and in energy 
between the two interfering neutrino mass-eigenstates;
\item
as $\sigma^2 \equiv \sigma_{\pi}^2 + \sigma_{\mu}^2 + \sigma_{\nu}^2$, 
$\sigma_{\pi}$, $\sigma_{\mu}$ and $\sigma_{\nu}$ being uncertainty 
associated respectively with $\vec{p}_{\pi}$, $\vec{p}_{\mu}$ and 
$\vec{p}_{\nu}$, becomes larger (smaller), the feature that each of 
the particles involved propagates along its classical trajectory 
(energies and momenta of the particles involved are conserved during 
the decay) becomes more prominent; and
\item
in the limit of $\sigma^2 \rightarrow \infty$, the oscillating terms 
would again be suppressed away.
\end{enumerate}

An approximate treatment which takes account of the two complementary 
features 
mentioned above is proposed and similarity and difference between our 
approach and that of Dolgov et al. are discussed.
\end{abstract}
\newpage
\def\thesection{\arabic{section}.}
\setcounter{footnote}{0}
\def\thefootnote{%
\arabic{footnote}}

\section{Introduction}
Neutrino oscillation\cite{1} is now one of the most exciting topics 
in particle physics, in non-accelerator as well as accelerator high 
energy physics and in astrophysics\cite{2}. There are now plenty of 
evidence in favor of presence of neutrino oscillations from 
atmospheric neutrino experiments, solar neutrino experiments, reactor 
neutrino experiments and accelerator neutrino experiments\cite{3}.

The standard formulas used to analyze the probability of neutrino 
oscillation between two flavor-eigenstates, e.g., $\nu_e$ and 
$\nu_{\mu}$, read
% (1.1)
\begin{eqnarray}
P_{\nu_\mu \rightarrow \nu_\mu} &=& P_{\nu_e \rightarrow \nu_e} 
~=~ 1-\sin^2 (2\theta) \sin^2(\frac{{m_1}^2-{m_2}^2}{4E_{\nu}}D), \no \\
P_{\nu_\mu \rightarrow \nu_e} &=& P_{\nu_e \rightarrow \nu_\mu} 
~=~ \sin^2 (2\theta) \sin^2(\frac{{m_1}^2-{m_2}^2}{4E_{\nu}}D),
\end{eqnarray}
where $E_{\nu}$ and $D$ are the energy and traveling distance of the 
neutrinos, $m_1$ and $m_2$ are the masses of two interfering 
mass-eigenstates, and $\theta$ is the mixing angle. 
% (1.2)
\begin{eqnarray}
l_{1 \leftrightarrow 2} ~\equiv~ \frac{4\pi E_{\nu}}{|m_1^2-m_2^2|}
\end{eqnarray}
is often referred to as the oscillation length. These formulas are 
usually derived on the basis of a plane-wave approach and without 
paying particular attention to how neutrinos are created/detected.

As is well known, the main physics involved in neutrino oscillation is 
quantum mechanics and this aspect of neutrino oscillation has widely 
been discussed in the literature\cite{4} since the pioneering work by 
Kayser\cite{5}. There are also not a few papers which treat neutrino 
oscillation field-theoretically\cite{6}. In particular, 
a field-theoretical approach of neutrino oscillation with both 
production process and detection process taken into account were 
recently developed by Giunti et al. in \cite{7} and by Asahara et al. 
in \cite{8}. In these works, the neutrino is treated as an intermediate 
particle, while all the external particles are treated as wave-packets.

We also have, in a series of papers\cite{9}$\sim$\cite{12}, developed 
wave-packet treatments of neutrino oscillation and addressed ourselves, 
in particular, to such questions as "How do neutrinos propagate?", 
"Equal energy or equal momentum or else?" and "How and why does the 
factor-of-two paradox arise?". In the present paper, after extending 
our previous treatment given in \cite{11} to a three-dimensional case, 
we shall go one step further to study pion decay 
$\pi \rightarrow \mu\nu$, with neutrino mixing taken into account and 
with emphasis placed on formal structure of its transition probability.

$\pi \to \mu \nu$ decay with neutrino oscillation taken into 
account has been investigated with emphasis placed on such a question 
as "Do muons oscillate?" by Dolgov et al. in \cite{13}. 
In their treatment, $(\vec{x}_0,t_0)$, space-time point where and 
when a pion decay occurs, is specified, each of the particles involved 
is supposed to propagate along its classical trajectory, and 
energy-momentum conservation is imposed by hand.\footnote
{
They, however, at the same time mention that, in quantum field theory, one 
includes integration over space-time point of decay in the definition 
of the relevant amplitude, and this integration leads to the 
conservation law.
}
 They first treat all the particles as plane-waves and then introduce 
momentum distribution for the pion alone.

The present work is more or less stimulated by the interesting work 
by Dolgov et al.\cite{13} and is organized as follows. In Sec.2 and 
in Appendix A, we focus on neutrino oscillation without paying 
attention to how the neutrino is created/detected. The neutrinos are 
treated as wave-packets and a number of comments related to those 
questions mentioned above are given. Pion decay with neutrino mixing 
taken into account is investigated in Sec.3 and Sec.4. In Sec.3, 
we define and derive the transition amplitude and probability for a 
sequence of processes $\pi^+ \to \mu^+ \nu_{\mu} \to \mu^+ 
\nu_{\alpha}$ with all the particles involved treated as wave-packets. 
The space-time point of creation of $\pi^+$ and of detection of $\mu^+$ 
and $\nu_{\alpha}$ is specified, while the space-time point of decay 
$(\vec{x}_0,t_0)$ is integrated out. The plane-wave limit and the total 
transition probability are also examined. In Sec.4, we propose an 
approximate treatment which takes account of the two complemetary 
features: energies and momenta are conserved on the one hand and each 
of the particles propagates along its classical trajectory on the other 
hand. Section 5 compares our approach with that of Dolgov et al. and 
Section 6 gives some concluding remarks. In Appendix B, some algebra 
relevant 
to Sec.4 is given.
\section{Three-dimensional wave-packet treatment of \\
neutrino oscillation}
Let $|\nu_{\alpha}\rangle$ ($\alpha = e,\mu, \cdots$) represent 
neutrino states associated with the electron, muon, $\cdots$, 
which are superpositions of the mass-eigenstates $|\nu_k\rangle$ 
having mass $m_k$ ($k = 1,2,\cdots$):\footnote
{
We shall assume $m_k \neq m_l$ for $k \neq l$. Also, since we are not 
interested in $CP$ or $T$ violation, we shall take $U = (U_{\alpha k})$ 
to be a real orthogonal matrix. In the two-generation case, 
$U_{\alpha k}$ may be expressed in terms of a single mixing angle 
$\theta$: $U_{e 1} = U_{\mu 2} = \cos\theta,~U_{e 2} = -U_{\mu 1} = 
\sin\theta$.
}
% (2.1)
\begin{eqnarray}
|\nu_\alpha\rangle = \sum_k U_{\alpha k}|\nu_k\rangle.
\end{eqnarray}

Suppose that a $\nu_{\mu}$ of momentum $\vec{p}_{\nu}$ with uncertainty 
(or dispersion) $\sigma_{\nu}$ is created at space-time point 
$(\vec{x}_0,t_0)$. Then, its state vector at $(\vec{x}_{\nu},t_{\nu})$ 
may be written as
% (2.2,2.3)
\begin{eqnarray}
&& |\nu_{\mu}(\vec{x}_{\nu},t_{\nu};\vec{x}_0,t_0;\vec{p}_{\nu},
\sigma_{\nu})\rangle ~=~ \sum_k U_{\mu k}|\nu_k
\rangle \phi_k(\vec{x}_{\nu 0},t_{\nu 0};\vec{p}_{\nu},\sigma_{\nu}), \\
&& \phi_k(\vec{x},t;\vec{p}_{\nu},\sigma_{\nu}) 
~=~ (\frac{\sigma_{\nu}}{\sqrt{\pi}})^{3/2}
\exp\{i(\vec{p}_{\nu}\vec{x}-E_k(\vec{p}_{\nu}) t) 
- \frac{1}{2}\sigma_{\nu}^2(\vec{x}-\vec{v}_k(\vec{p}_{\nu}) t)^2\}, 
\no \\
\end{eqnarray}
where
\begin{eqnarray*}
&& \vec{x}_{\nu 0} ~=~ \vec{x}_{\nu}-\vec{x}_0, \qquad 
t_{\nu 0} ~=~ t_{\nu}-t_0, \\
&& E_k(\vec{p}) ~=~ \sqrt{\vec{p}^2+m_k^2}, \qquad
\vec{v}_k(\vec{p}) ~=~ \frac{dE_k(\vec{p})}{d\vec{p}} ~=~ \frac{\vec{p}}{E_k(\vec{p})}.
\end{eqnarray*}

As is well known\cite{14}, the wave function of the form given by 
Eq.(2.3) follows readily, if one
\begin{enumerate}
\item
starts from a superposition of plane-wave functions:\footnote
{Neglecting spin degree of freedom, we shall assume that the neutrinos 
(and the charged leptons as well) obey the Klein-Gordon equation.
}
% (2.4)
\begin{eqnarray}
\phi_k(\vec{x},t;\vec{p}_{\nu},\sigma_{\nu}) 
~=~ \int_{-\infty}^{\infty} \frac{d^3p}{(2\pi)^{3/2}} 
f(\vec{p};\vec{p}_{\nu},\sigma_{\nu}) 
\exp\{i(\vec{p}\vec{x}-E_k(\vec{p})t)\};
\end{eqnarray}
\item
takes, as the momentum distribution function 
$f(\vec{p};\vec{p}_{\nu},\sigma_{\nu})$,
% (2.5)
\begin{eqnarray}
f(\vec{p};\vec{p}_{\nu},\sigma_{\nu}) 
~=~ \frac{1}{(\sqrt{\pi}\sigma_{\nu})^{3/2}} 
\exp\{-\frac{(\vec{p}-\vec{p}_{\nu})^2}{2\sigma_{\nu}^2}\};
\end{eqnarray}
\item
expands $E_k(\vec{p})$ around $\vec{p}_{\nu}$ as
% (2.6)
\begin{eqnarray}
E_k(\vec{p}) ~=~ E_k(\vec{p}_{\nu})+\vec{v}_k(\vec{p}_{\nu})
(\vec{p}-\vec{p}_{\nu});
\end{eqnarray}
\item
and performs the $\vec{p}$-integration involved in Eq.(2.4) explicitly.
\end{enumerate}
Note that we have normalized $f(\vec{p};\vec{p}_{\nu},\sigma_{\nu})$ 
and $\phi_k(\vec{x},t;\vec{p}_{\nu},\sigma_{\nu})$ as
% (2.7)
\begin{eqnarray}
\int_{-\infty}^{\infty}d^3p~|f(\vec{p};\vec{p}_{\nu},\sigma_{\nu})|^2 
~=~ \int_{-\infty}^{\infty}d^3x~
|\phi_k(\vec{x},t;\vec{p}_{\nu},\sigma_{\nu})|^2 ~=~ 1,
\end{eqnarray}
and that, to keep this normalization condition valid, we shall define 
and express the plane-wave limit of 
$\phi_k(\vec{x},t;\vec{p}_{\nu},\sigma_{\nu})$ as
% (2.8)
\begin{eqnarray}
\lim_{\sigma_{\nu} \To 0} \phi_k(\vec{x},t;\vec{p}_{\nu},\sigma_{\nu}) 
~=~ \frac{1}{V^{1/2}} \exp\{i(\vec{p}_{\nu}\vec{x}-E_k(\vec{p}_{\nu})t)\},
\end{eqnarray}
that is, letting $\sigma_{\nu} \to 0$ in the exponent and, at the same 
time, on confining the neutrino plane-waves within a spatial cube of 
volume $V$, letting the normalization constant 
$N_{\nu} \equiv (\sigma_{\nu}/\sqrt{\pi})^{3/2} \to 1/V^{1/2}$.\footnote
{
From the uncertainty relation, one may interpret $V_{\rm wave-packet} 
\equiv (\sqrt{\pi}/\sigma_{\nu})^3$ as the volume of a spatial region 
within which the neutrino wave-packets are appreciable and express 
$N_{\nu} \to 1/V^{1/2}$ as $V_{\rm wave-packet} \to V$.
} 

The amplitude for a $\nu_{\mu}$ created at $(\vec{x}_0,t_0)$ to be 
detected as a $\nu_{\alpha}$ at $(\vec{x}_{\nu},t_{\nu})$ is calculated 
as
% (2.9)
\begin{eqnarray}
\lefteqn{A_{\nu_{\mu} \rightarrow \nu_{\alpha}}(\vec{x}_{\nu 0},t_{\nu 0})
~=~ \langle \nu_{\alpha}|\nu_{\mu}(\vec{x}_{\nu},t_{\nu};\vec{x}_0,t_0;\vec{p}_{\nu},\sigma_{\nu}) \rangle } \hspace{1cm} \no \\
&=& \sum_{k} U_{\mu k} U_{\alpha k}~\phi_k(\vec{x}_{\nu 0},t_{\nu 0};\vec{p}_{\nu},\sigma_{\nu}) \no \\
&=& N_{\nu}~\sum_{k} U_{\mu k} U_{\alpha k}~\exp\{i(\vec{p}_{\nu}\vec{x}_{\nu 0}-E_k(\vec{p}_{\nu}) t_{\nu 0}) - \frac{1}{2}\sigma_{\nu}^2(\vec{x}_{\nu 0}-\vec{v}_k(\vec{p}_{\nu}) t_{\nu 0})^2\}, \no \\
\end{eqnarray}
and the corresponding propability is calculated as
% (2.10)
\begin{eqnarray}
\lefteqn{P_{\nu_{\mu} \rightarrow \nu_{\alpha}}(\vec{x}_{\nu 0},t_{\nu 0})
~=~ |A_{\nu_{\mu} \rightarrow \nu_{\alpha}}(\vec{x}_{\nu 0},t_{\nu 0})|^2} \hspace{1cm} \no \\
&=& \sum_{k,l} U_{\mu k} U_{\alpha k} U_{\mu l} U_{\alpha l}~
   \phi_k(\vec{x}_{\nu 0},t_{\nu 0};\vec{p}_{\nu},\sigma_{\nu})
   \phi_l^*(\vec{x}_{\nu 0},t_{\nu 0};\vec{p}_{\nu},\sigma_{\nu}) \no \\
&=& N_{\nu}^2 \sum_{k,l} U_{\mu k} U_{\alpha k} U_{\mu l} U_{\alpha l} 
\exp\{-iE_{[kl]}(\vec{p}_{\nu})t_{\nu 0}\} \no \\
&& \exp\{-\frac{1}{2}\sigma_{\nu}^2(\vec{x}_{\nu 0}-\vec{v}_kt_{\nu 0})^2-\frac{1}{2}\sigma_{\nu}^2(\vec{x}_{\nu 0}-\vec{v}_lt_{\nu 0})^2\} \no \\
&=& N_{\nu}^2 \sum_{k,l} U_{\mu k} U_{\alpha k} U_{\mu l} U_{\alpha l} \cos(E_{[kl]}(\vec{p}_{\nu})t_{\nu 0}) \no \\
&& \exp\{-\sigma_{\nu}^2(\vec{x}_{\nu 0}-\vec{v}_{(kl)}(\vec{p}_{\nu})t_{\nu 0})^2\} 
   \exp\{-\frac{1}{4}\sigma_{\nu}^2(\vec{v}_{[kl]}(\vec{p}_{\nu}))^2t_{\nu 0}^2\},
\end{eqnarray}
where
\begin{eqnarray*}
E_{(kl)}(\vec{p}_{\nu}) &=& \frac{1}{2}(E_k(\vec{p}_{\nu})+E_l(\vec{p}_{\nu})),
 \qquad E_{[kl]}(\vec{p}_{\nu}) ~=~ E_{k}(\vec{p}_{\nu})-E_l(\vec{p}_{\nu}), \\
\vec{v}_{(kl)}(\vec{p}_{\nu}) &=& \frac{1}{2}(\vec{v}_k(\vec{p}_{\nu})+\vec{v}_l(\vec{p}_{\nu})), \qquad \vec{v}_{[kl]}(\vec{p}_{\nu}) ~=~ \vec{v}_k(\vec{p}_{\nu})-\vec{v}_l(\vec{p}_{\nu}).
\end{eqnarray*}
With Eq.(2.7), we have
%(2.11)
\begin{eqnarray}
\sum_{\alpha} \int d^3x_{\nu}~P_{\nu_{\mu} \rightarrow \nu_{\alpha}}(\vec{x}_{\nu 0},t_{\nu 0}) ~=~ 1.
\end{eqnarray}

A couple of remarks are in order as regards Eq.(2.10).
\begin{enumerate}
\item
Each term in Eq.(2.10) contains explicitly a factor which implies that 
the space-time point $(\vec{x}_{\nu},t_{\nu})$ where and when the 
neutrinos are detected nearly satisfies
% (2.12)
\begin{equation}
\vec{x}_{\nu} ~=~ \vec{x}_0 + \vec{v}_{(kl)}(\vec{p}_{\nu})(t_{\nu}-t_0),
\end{equation}
which is precisely what could be referred to as classical trajectory 
of the neutrinos. We may therefore, assuming that 
$(\vec{x}_{\nu 0}-\vec{v}_{(kl)}
(\vec{p}_{\nu})t_{\nu 0})^2 \ll \sigma_{\nu}^{-2}$ 
is satisfied for any $k$ and $l$, express Eq.(2.10) approximately as
% (2.13)
\begin{eqnarray}
\lefteqn{P_{\nu_{\mu} \rightarrow \nu_{\alpha}}(\vec{x}_{\nu 0},t_{\nu 0})} \hspace{1cm} \no \\
&\simeq& N_{\nu}^2 \sum_{k,l} U_{\mu k} U_{\alpha k} U_{\mu l} U_{\alpha l} \cos(E_{[kl]}(\vec{p}_{\nu})t_{\nu 0}) \exp\{-\frac{1}{4}\sigma_{\nu}^2(\vec{v}_{[kl]}(\vec{p}_{\nu}))^2t_{\nu 0}^2\} \no \\
&=& N_{\nu}^2 \sum_{k,l} U_{\mu k} U_{\alpha k} U_{\mu l} U_{\alpha l} 
\cos(2\pi\frac{D}{l_{k \leftrightarrow l}}) \exp\{-\frac{1}{4}\sigma_{\nu}^2\frac{(\vec{v}_{[kl]}(\vec{p}_{\nu}))^2}{(\vec{v}_{(kl)}(\vec{p}_{\nu}))^2}D^2\}, \no \\
\end{eqnarray}
where $D \equiv |\vec{x}_{\nu 0}| = |\vec{v}_{(kl)}(\vec{p}_{\nu})|
t_{\nu 0}$ is the traveling distance of the neutrinos, and
% (2.14)
\begin{eqnarray}
l_{k \to l} ~=~ \frac{2\pi|\vec{v}_{(kl)}(\vec{p}_{\nu})|}{|E_{[kl]}(\vec{p}_{\nu})|} 
~=~ \frac{4\pi E_{(kl)}(\vec{p}_{\nu})|\vec{v}_{(kl)}(\vec{p}_{\nu})|}{|m_k^2-m_l^2|}.
\end{eqnarray}
\item
In each of interference terms, the oscillating factor 
$\cos(E_{[kl]}(\vec{p}_{\nu})t_{\nu 0})$ is multiplied by a factor, denoted by 
$\xi_{kl}$, which implies that this term is significant only when the 
condition
% (2.17)
\begin{eqnarray}
t_{\nu}-t_0 ~\lesssim~ \frac{2}{\sigma_{\nu}|\vec{v}_{[kl]}(\vec{p}_{\nu})|} 
\qquad {\rm or} \qquad D ~\lesssim~ \frac{2|\vec{v}_{(kl)}(\vec{p}_{\nu})|}{\sigma_{\nu}|\vec{v}_{[kl]}(\vec{p}_{\nu})|},
\end{eqnarray}
is satisfied. This effect, often referred to as coherent condition for 
neutrino oscillation\cite{15}, manifests itself explicitly in our 
wave-packet treatment but not in conventional plane-wave treatments. 
Note also that $\vec{v}_k(\vec{p}_{\nu}) \neq \vec{v}_l(\vec{p}_{\nu})$ 
for $k \neq l$ only results in appearence of this effect, but by no 
means affects the oscillating factor and hence the oscillation period 
$2\pi/E_{[kl]}(\vec{p}_{\nu})$ or oscillation length 
$l_{k \to l}$.\footnote
{
If, noting that the magnitude of the amplitude 
$A_{\nu_{\mu} \rightarrow \nu_{\alpha}}(\vec{x}_{\nu 0},t_{\nu 0})$, 
Eq.(2.9), has a peak at $\vec{x}_{\nu 0}=\vec{v}_k(\vec{p}_{\nu})
t_{\nu 0}$, one substitutes this into its phase factor to obtain 
$\exp\{-i(m_k^2/E_k(\vec{p}_{\nu}))t_{\nu 0}\}$, 
the phase factor of the probability (2.10) would become 
$\exp\{-i(m_k^2/E_k(\vec{p}_{\nu})-m_l^2/E_l(\vec{p}_{\nu}))t_{\nu 0}\}$, 
which will results in an oscillation period nearly twice as large as the 
standard one for relativistic neutrinos. This is where the so-called 
"factor-of-two paradox"\cite{16} arises and is obviously an inadequate 
prescription.
} 
\item
In conventional plane-wave approaches, some prefer to work with 
an "equal-momentum prescription" and others an "equal-energy 
prescription"\cite{5},\cite{16}. We like to point out that the 
wave-packet treatment given here reduces smoothly, in the plane-wave 
limit, to the former prescription\footnote
{
We shall give a more general three-dimensional wave-packet treatment of 
neutrino oscillation in Appendix A. Note that, for relativistic 
neutrinos and in the two-generation case, Equations (2.13) and (2.14) 
coincide (apart from the normalization factor) with Eqs.(1.1) and 
(1.2), if one let $\sigma_{\nu} \To 0$ and identifies 
$E_{(kl)}(\vec{p}_{\nu})$ with $E_{\nu}$.
}
 and that, although these two prescriptions give approximately the same 
oscillation formulas for relativistic neutrinos, they have to be 
distinguished from each other in general and conceptually\cite{10}.
\end{enumerate}
\section{Wave-packet treatment of $\pi \to \mu \nu$ Decay}
\subsection*{3.1. $\pi \to \mu \nu_{\mu}$ decay followed by 
$\nu_{\mu} \to \nu_{\alpha}$ transition}
We now attempt to formulate such a sequence of processes: a $\pi^+$ 
produced at $(\vec{x}_{\pi},t_{\pi})$ with momentum $\vec{p}_{\pi}$ 
decays into a $\mu^+$ and a $\nu_{\mu}$ somewhere in space-time and 
then the $\mu^+$ is detected at $(\vec{x}_{\mu},t_{\mu})$ with momentum 
$\vec{p}_{\mu}$ and a $\nu_{\alpha}$ is detected at 
$(\vec{x}_{\nu},t_{\nu})$ with momentum $\vec{p}_{\nu}$. Taking into 
account that the momenta relevant may not be free from some uncertainty 
or dispersion, we shall treat all the particles involved as wave-packets 
and examine the plane-wave limit later on.

If the decay space-time point is denoted by $(\vec{x}_0,t_0)$, we may 
use the amplitude we have introduced, Eq.(2.9), to describe a 
$\nu_{\mu}$ created at the decay point to be detected as a 
$\nu_{\alpha}$ at $(\vec{x}_{\nu},t_{\nu})$. Similarly, the amplitude 
for a $\mu^+$ created at the decay point to propagate to its detection 
point $(\vec{x}_{\mu},t_{\mu})$ and the amplitude for a $\pi^+$ 
produced at $(\vec{x}_{\pi},t_{\pi})$ to propagate to its decay point 
may be described respectively by\footnote
{
From now on, momentum dependence of energies and velocities will not be 
indicated explicitly, if not particularly necessary.
}
%
% (3.1)
\begin{eqnarray}
A_{\mu}(\vec{x}_{\mu 0},t_{\mu 0}) 
&=& \phi_{\mu}(\vec{x}_{\mu 0},t_{\mu 0};\vec{p}_{\mu},\sigma_{\mu}) \no \\
&=& \int \frac{d^3p}{(2\pi)^{3/2}} f(\vec{p};\vec{p}_{\mu},\sigma_{\mu}) 
    \exp\{i(\vec{p}\vec{x}_{\mu 0}-E_{\mu}(\vec{p})t_{\mu 0})\} \no \\
&=& N_{\mu} \exp\{i(\vec{p}_{\mu}\vec{x}_{\mu 0}-E_{\mu}t_{\mu 0})-\frac{1}{2}\sigma_{\mu}^2(\vec{x}_{\mu 0}-\vec{v}_{\mu}t_{\mu 0})^2\}, \no \\
A_{\pi}(\vec{x}_{0 \pi},t_{0 \pi})
&=& \phi_{\pi}(\vec{x}_{0 \pi},t_{0 \pi};\vec{p}_{\pi},\sigma_{\pi}) 
~=~ \phi_{\pi}^*(\vec{x}_{\pi 0},t_{\pi 0};\vec{p}_{\pi},\sigma_{\pi}) \no \\
&=& \int \frac{d^3p}{(2\pi)^{3/2}} f(\vec{p};\vec{p}_{\pi},\sigma_{\pi}) \exp\{-i(\vec{p}\vec{x}_{\pi 0}-E_{\pi}(\vec{p})t_{\pi 0})\} \no \\
&=& N_{\pi} \exp\{-i(\vec{p}_{\pi}\vec{x}_{\pi 0}-E_{\pi}t_{\pi 0})-\frac{1}{2}\sigma_{\pi}^2(\vec{x}_{\pi 0}-\vec{v}_{\pi}t_{\pi 0})^2\},
\end{eqnarray}
where $\sigma_{\mu}$ and $\sigma_{\pi}$ are uncertainties associated 
respectively with $\vec{p}_{\mu}$ and $\vec{p}_{\pi}$ and 
\begin{eqnarray*}
&& N_{\lambda} ~=~ (\frac{\sigma_{\lambda}}{\sqrt{\pi}})^{3/2}, \qquad 
\vec{x}_{\lambda 0} ~=~ \vec{x}_{\lambda}-\vec{x}_0, \qquad t_{\lambda 0} ~=~ t_{\lambda}-t_0, \\
&& E_{\lambda} ~=~ \sqrt{\vec{p}_{\lambda}^2+m_{\lambda}^2}, \qquad 
\vec{v}_{\lambda} ~=~ \frac{\vec{p}_{\lambda}}{E_{\lambda}}, \qquad \lambda~=~\mu,~\pi.
\end{eqnarray*}

The transition amplitude for 
$\pi^+ \rightarrow \mu^+ \nu_{\mu} \rightarrow \mu^+ \nu_{\alpha}$ 
with the decay space-time point $(\vec{x}_0,t_0)$ specified is then 
given by
% (3.2)
\begin{eqnarray}
A_{\pi \rightarrow \mu \nu_{\mu} \rightarrow \mu \nu_{\alpha}}(\vec{x}_0,t_0) 
&=& A_{\nu_{\mu} \to \nu_{\alpha}}(\vec{x}_{\nu 0},t_{\nu 0}) A_{\mu}(\vec{x}_{\mu 0},t_{\mu 0}) A_{\pi}(\vec{x}_{0 \pi},t_{0 \pi}) \no \\
&=& \sum_k U_{\mu k}U_{\alpha k} A_k(\vec{x}_0,t_0),
\end{eqnarray}
where
% (3.3)
\begin{eqnarray}
A_k(\vec{x}_0,t_0) 
&=& \phi_k(\vec{x}_{\nu 0},t_{\nu 0};\vec{p}_{\nu},\sigma_{\nu})
    \phi_{\mu}(\vec{x}_{\mu 0},t_{\mu 0};\vec{p}_{\mu},\sigma_{\mu}) 
    \phi_{\pi}^*(\vec{x}_{\pi 0},t_{\pi 0};\vec{p}_{\pi},\sigma_{\pi}) \no \\
&=& N_{\nu}\exp\{i(\vec{p}_{\nu}\vec{x}_{\nu 0}-E_kt_{\nu 0}) 
   -\frac{1}{2}\sigma_{\nu}^2(\vec{x}_{\nu 0}-\vec{v}_kt_{\nu 0})^2\}, \no \\
&&  N_{\mu}\exp\{i(\vec{p}_{\mu}\vec{x}_{\mu 0}-E_{\mu}t_{\mu 0}) 
   -\frac{1}{2}\sigma_{\mu}^2(\vec{x}_{\mu 0}-\vec{v}_{\mu}t_{\mu 0})^2\} \no \\
&&  N_{\pi}\exp\{-i(\vec{p}_{\pi}\vec{x}_{\pi 0}-E_{\pi}t_{\pi 0}) 
   -\frac{1}{2}\sigma_{\pi}^2(\vec{x}_{\pi 0}-\vec{v}_{\pi}t_{\pi 0})^2 \}.
\end{eqnarray}

The transition amplitude for the whole process we have specified in the 
begining of this subsection may be obtained by integrating Eq.(3.2) 
with respect to $\vec{x}_0$ and $t_0$,\footnote
{
It is understood that the $t_0$-integration extends over a infinitely 
large time interval $T$.
}
%
% (3.4)
\begin{eqnarray}
A_{\pi \rightarrow \mu \nu_{\mu} \rightarrow \mu \nu_{\alpha}} 
&=& \int d^3x_0 \int dt_0~A_{\pi \rightarrow \mu \nu_{\mu} \rightarrow \mu \nu_{\alpha}} (\vec{x}_0,t_0) \no \\
&=& \sum_k U_{\mu k}U_{\alpha k} A_k,
\end{eqnarray}
and the corresponding transition probability per unit time interval is 
given by
% (3.5)
\begin{eqnarray}
P_{\pi \rightarrow \mu \nu_{\mu} \rightarrow \mu \nu_{\alpha}} 
&=& \frac{1}{T}~|A_{\pi \rightarrow \mu \nu_{\mu} \rightarrow \mu \nu_{\alpha}}|^2 \no \\
&=& \sum_{k,l}U_{\mu k}U_{\alpha k}U_{\mu l}U_{\alpha l}~G_{kl},
\end{eqnarray}
where
% (3.6,3.7)
\begin{eqnarray}
A_k &=& \int d^3x_0 \int dt_0~A_k(\vec{x}_0,t_0), \\
G_{kl} &=& \frac{1}{T}~A_kA_l^* ~=~ \frac{1}{T}~\int d\vec{x}_0 \int dt_0 \int d\vec{x}_0' \int dt_0'~A_k(\vec{x}_0,t_0)A_l^*(\vec{x}_0',t_0'). \no \\
\end{eqnarray}
\subsection*{3.2. Transition amplitude and probability}
To calculate $A_k$, Eq.(3.6), we write Eq.(3.3) as
% (3.8)
\begin{eqnarray}
A_k(\vec{x}_0,t_0) 
&=& N \exp\{i(\theta_k-\Delta \vec{p}\vec{x}_0+\Delta E_kt_0)-\frac{1}{2}\sigma^2y_k(\vec{x}_0,t_0)\},
\end{eqnarray}
where
% (3.9)
\begin{eqnarray}
N &=& N_{\nu}N_{\mu}N_{\pi} ~=~ (\frac{\sigma_{\nu}\sigma_{\mu}\sigma_{\pi}}{\pi\sqrt{\pi}})^{3/2}, \qquad \sigma^2 ~=~ \sigma_{\nu}^2+\sigma_{\mu}^2+\sigma_{\pi}^2, \no \\
\Delta\vec{p} &=& \vec{p}_{\nu}+\vec{p}_{\mu}-\vec{p}_{\pi}, \qquad 
\Delta E_k ~=~ E_k+E_{\mu}-E_{\pi}, \no \\
\theta_k &=&~ \vec{p}_{\nu}\vec{x}_{\nu}-E_kt_{\nu} 
        +\vec{p}_{\mu}\vec{x}_{\mu}-E_{\mu}t_{\mu} 
        -\vec{p}_{\pi}\vec{x}_{\pi}+E_{\pi}t_{\pi}, \no \\
y_k(\vec{x}_0,t_0) 
&=& \frac{1}{\sigma^2}\{\sigma_{\nu}^2(\vec{x}_{\nu 0}-\vec{v}_k t_{\nu 0})^2 
+ \sigma_{\mu}^2(\vec{x}_{\mu 0}-\vec{v}_{\mu} t_{\mu 0})^2 
+ \sigma_{\pi}^2(\vec{x}_{\pi 0}-\vec{v}_{\pi} t_{\pi 0})^2\}. \no \\
\end{eqnarray}
Introducing\footnote
{
$\bra\vec{u},\vec{w}\ket_k$ may be expressed also as
% (3.10)
\begin{eqnarray}
\bra\vec{u},\vec{w}\ket_k 
&=& \frac{1}{\sigma^4}\{\sigma_{\nu}^2 \sigma_{\mu}^2 \vec{u}_{[k\mu]}\vec{w}_{[k\mu]} + \sigma_{\nu}^2 \sigma_{\pi}^2 \vec{u}_{[k\pi]}\vec{w}_{[k\pi]} +  \sigma_{\mu}^2 \sigma_{\pi}^2 \vec{u}_{[\mu\pi]}\vec{w}_{[\mu\pi]}\},
\end{eqnarray}
where
\begin{eqnarray*}
\vec{u}_{[\kappa\lambda]} ~=~ \vec{u}_{\kappa}-\vec{u}_{\lambda}, \qquad \qquad \kappa,~\lambda ~=~ k,~\mu,~\pi.
\end{eqnarray*}
}
% (3.11)
\begin{eqnarray}
\vec{X}_k &=& \vec{x}_{\nu}-\vec{v}_k t_{\nu}, \qquad \vec{X}_{\mu} ~=~ \vec{x}_{\mu}-\vec{v}_{\mu}t_{\mu}, \qquad \vec{X}_{\pi} ~=~ \vec{x}_{\pi}-\vec{v}_{\pi}t_{\pi}, \no \\
\bra\vec{u}\ket_k &=& (\sigma_{\nu}^2\vec{u}_k+\sigma_{\mu}^2\vec{u}_{\mu}+\sigma_{\pi}^2\vec{u}_{\pi})/\sigma^2, \no \\
\bra\vec{u}\vec{w}\ket_k &=& (\sigma_{\nu}^2\vec{u}_k\vec{w}_k+\sigma_{\mu}^2\vec{u}_{\mu}\vec{w}_{\mu}+\sigma_{\pi}^2\vec{u}_{\pi}\vec{w}_{\pi})/\sigma^2, \no \\
\bra\vec{u},\vec{w}\ket_k &=& \bra\vec{u}\vec{w}\ket_k-\bra\vec{u}\ket_k\bra\vec{w}\ket_k, \qquad \qquad \vec{u},~\vec{w}=\vec{v},~\vec{X}, 
\end{eqnarray}
we further express $y_k(\vec{x}_0,t_0)$ as
% (3.12)
\begin{eqnarray}
y_k(\vec{x}_0,t_0) &=& (\vec{x}_0-\bra\vec{v}\ket_kt_0-\bra\vec{X}\ket_k)^2+a_k(t_0-t_k)^2+c_k-a_kt_k^2,
\end{eqnarray}
and $i(\theta_k-\Delta\vec{p}\vec{x}_0+\Delta E_k t_0)
-\frac{1}{2}\sigma^2y_k(\vec{x}_0,t_0)$ as
% (3.13)
\begin{eqnarray}
\lefteqn{i(\theta_k-\Delta\vec{p}\vec{x}_0+\Delta E_k t_0)-\frac{1}{2}\sigma^2y_k(\vec{x}_0,t_0)} \hspace{0.5cm} \no \\
&=& i(\theta_k-\Delta\vec{p}\bra\vec{X}\ket_k+\Delta\tilde{E}_kt_k) \no \\
&&  -\frac{1}{2}\sigma^2(c_k -a_kt_k^2)-\frac{1}{2\sigma^2}(\Delta\vec{p})^2-\frac{1}{2\sigma^2a_k}(\Delta\tilde{E}_k)^2 \no \\
&& -\frac{1}{2}\sigma^2(\vec{x}_0-\bra\vec{v}\ket_kt_0-\bra\vec{X}\ket_k+i\frac{1}{\sigma^2}\Delta\vec{p})^2 
 -\frac{1}{2}\sigma^2a_k(t_0-t_k-i\frac{1}{\sigma^2a_k}\Delta\tilde{E}_k)^2, \no \\
\end{eqnarray}
where
% (3.14)
\begin{eqnarray}
a_k &=& \bra\vec{v},\vec{v}\ket_k, \qquad b_k 
~=~ \bra\vec{v},\vec{X}\ket_k, \qquad 
c_k ~=~ \bra\vec{X},\vec{X}\ket_k, \no \\
t_k &=& -b_k/a_k, \qquad \Delta\tilde{E}_k 
~=~ \Delta E_k-\bra\vec{v}\ket_k\Delta\vec{p}.
\end{eqnarray}

Substituting Eqs.(3.8) and (3.13) into Eq.(3.6) and performing the 
$\vec{x}_0$- and $t_0$-integrations over the whole space-time, 
we find
% (3.15)
\begin{eqnarray}
A_k &=& N (\frac{2\pi}{\sigma^2})^{3/2} (\frac{2\pi}{\sigma^2a_k})^{1/2} \exp\{i(\theta_k - \Delta\vec{p}\bra\vec{X}\ket_k + \Delta\tilde{E}_kt_k)\} \no \\
&& \exp\{-\frac{1}{2\sigma^2}(\Delta\vec{p})^2-\frac{1}{2\sigma^2a_k}(\Delta\tilde{E}_k)^2-\frac{1}{2}\sigma^2(c_k-a_kt_k^2)\}.
\end{eqnarray}
Substituting Eq.(3.15) into Eq.(3.7), we obtain\footnote
{
One may substitute Eqs.(3.8) and (3.13) directly into Eq.(3.7) and 
perform the $\vec{x}_0$-, $t_0$-, $\vec{x}_0'$- and $t_0'$-integrations, 
by changing the integration variables from $\vec{x}_0$ and 
$\vec{x}_0'$ to $\vec{x}_- \equiv \vec{x}_0-\vec{x}_0$ and 
$\vec{x}_+ \equiv (\vec{x}_0+\vec{x}_0')/2$, 
and from $t_0$ and $t_0'$ to $t_- \equiv t_0-t_o'$ and 
$t_+ \equiv (a_kt_0+a_lt_0')/(a_k+a_l)$, to obtain the same result.
}
% (3.16)
\begin{eqnarray}
G_{kl} &=& \frac{1}{T}~A_kA_l^* \nonumber \\
&=& \frac{1}{T}~N~(\frac{2\pi}{\sigma^2})^{3/2}~(\frac{2\pi}{\sigma^2a_k})^{1/2}~\exp\{i(\theta_k - \Delta\vec{p}\bra\vec{X}\ket_k + \Delta\tilde{E}_kt_k)\} \nonumber \\
&& \exp\{-\frac{1}{2\sigma^2}(\Delta\vec{p})^2-\frac{1}{2\sigma^2a_k}(\Delta\tilde{E}_k)^2-\frac{1}{2}\sigma^2(c_k-a_kt_k^2)\} \nonumber \\
&& N~(\frac{2\pi}{\sigma^2})^{3/2}~(\frac{2\pi}{\sigma^2a_l})^{1/2}~\exp\{-i(\theta_l - \Delta\vec{p}\bra\vec{X}\ket_l + \Delta\tilde{E}_lt_l)\} \nonumber \\
&& \exp\{-\frac{1}{2\sigma^2}(\Delta\vec{p})^2-\frac{1}{2\sigma^2a_l}(\Delta\tilde{E}_l)^2-\frac{1}{2}\sigma^2(c_l-a_lt_l^2)\} \no \\
&=& N_{kl}~\exp\{-\sigma^2Z_{kl}-\frac{1}{\sigma^2}H_{kl}-i\Theta_{kl}\},
\end{eqnarray}
where
% (3.17,3.18,3.19,3.20)
\begin{eqnarray}
N_{kl} &=& \frac{1}{T}~N^2~(\frac{2\pi}{\sigma^2})^4~(\frac{1}{a_ka_l})^{1/2}, \\
Z_{kl} &=& \frac{1}{2}(c_k-a_kt_k^2+c_l-a_lt_l^2), \\
H_{kl} &=& (\Delta\vec{p})^2+\frac{1}{2a_k}(\Delta\tilde{E}_k)^2+
\frac{1}{2a_l}(\Delta\tilde{E}_l)^2, \\
\Theta_{kl} &=& -(\theta_k - \Delta\vec{p}\bra\vec{X}\ket_k + \Delta\tilde{E}_kt_k) + (\theta_l - \Delta\vec{p}\bra\vec{X}\ket_l + \Delta\tilde{E}_lt_l) \no \\
&=& E_{[kl]}t_{\nu}+\Delta\vec{p}(\bra\vec{X}\ket_k-\bra\vec{X}\ket_l)-(\Delta\tilde{E}_kt_k-\Delta\tilde{E}_lt_l).
\end{eqnarray}

A couple of remarks as regards ''trajectories of particles'' are 
in order.
\begin{enumerate}
\item
From Eq.(3.8), one sees that $|A_k(\vec{x}_0,t_0)|$ has a peak at 
$y_k(\vec{x}_0,t_0)= 0$, which, from Eqs.(3.9) and (3.12), implies 
on the one hand that
% (3.21)
\begin{eqnarray}
\vec{x}_{\nu 0}-\vec{v}_k t_{\nu 0} &=& 0, \qquad 
\vec{x}_{\mu 0}-\vec{v}_{\mu} t_{\mu 0} ~=~ 0, \qquad
\vec{x}_{\pi 0}-\vec{v}_{\pi} t_{\pi 0} ~=~ 0,
\end{eqnarray}
and on the other hand that
% (3.22)
\begin{eqnarray}
\vec{x}_0-\bra\vec{v}\ket_kt_0-\bra\vec{X}\ket_k ~=~ 0, \qquad
t_0-t_k ~=~ 0, \qquad
c_k-a_kt_k^2 ~=~ 0.
\end{eqnarray}
The three equations in Eq.(3.22) may be derived somehow directly as 
some weighted averages of the three equations in Eq.(3.21).\footnote
{Note that the three equations in Eq.(3.21) may be rewritten as
% (3.23)
\begin{eqnarray}
\vec{x}_0-\vec{v}_kt_0 ~=~ \vec{X}_k, \qquad \vec{x}_0-\vec{v}_{\mu}t_0 ~=~ \vec{X}_{\mu}, \qquad \vec{x}_0-\vec{v}_{\pi}t_0 ~=~ \vec{X}_{\pi},
\end{eqnarray}
from which it follows that
% (3.24)
\begin{eqnarray}
\vec{v}_{[k\mu]}t_0 &=& -\vec{X}_{[k\mu]}, \qquad 
\vec{v}_{[\mu\pi]}t_0 ~=~ -\vec{X}_{[\mu\pi]}, \qquad 
\vec{v}_{[\pi k]}t_0 ~=~ -\vec{X}_{[\pi k]}.
\end{eqnarray}
The first equation in Eq.(3.22) follows from Eq.(3.23), while the 
second and third equations in Eq.(3.22) follow from Eq.(3.24).
}
\item
From Eqs.(3.15) and (3.16), one sees that $|A_k|$, as well as $G_{kk}$, 
has a peak at $c_k-a_kt_k^2=0$. Thus, even after the integrations with 
respect to $\vec{x}_0$ and $t_0$ have been carried out, the third 
equation in Eq.(3.22) remains meaningful and may be regarded as 
something which reflects that each of the particles involved in the 
decay propagates along its classical trajectory, Eq.(3.21). 
\item
For $k \neq l$, $c_k-a_kt_k^2=0$ and $c_l-a_lt_l^2=0$ are not compatible 
with each other and $Z_{kl} \equiv (c_k-a_kt_k^2+c_l-a_lt_l^2)/2 = 0$ 
never holds. Thus, for the interference terms, $G_{kl}$ with $k \neq l$, 
it is actually not quite clear as to if and how trajectories may be 
defined for the particles involved in the decay. We shall come back to 
this question in the next section.
\end{enumerate}
\subsection*{3.3. Plane-wave limit}
Let us examine the plane-wave limit: $\sigma \to 0$ or, eqivalently, 
$\sigma_{\nu}$, $\sigma_{\mu}$, $\sigma_{\pi}$ $\to$ 0. By going back 
to Eqs.(3.3) and (3.6) and trivially performing integrations with 
respect to $\vec{x}_0$ and $t_0$, one finds
\begin{eqnarray*}
\lim_{\sigma_{\pi},\sigma_{\mu},\sigma_{\nu} \To 0}~A_k &=& \frac{1}{V^{3/2}}~(2\pi)^4~\exp\{i\theta_k\}~\delta^3(\Delta\vec{p})~\delta(\Delta E_k), \no \\
\end{eqnarray*}
from which it follows that\footnote
{
To derive Eq.(3.25), it is taken into account that, for $k \neq l$, 
$\Delta E_k=0$ and $\Delta E_l=0$ are not compatible with each other 
and use is made of the farmiliar technique to handle with the square of 
a $\delta$-function\cite{17}.
}
% (3.25)
\begin{eqnarray}
\lim_{\sigma_{\pi},\sigma_{\mu},\sigma_{\nu} \To 0}~G_{kl} 
&=& \frac{1}{T}~\frac{1}{V^3}~(2\pi)^8~\exp\{-iE_{[kl]}t_{\nu}\}~[\delta^3(\Delta\vec{p})]^2~\delta(\Delta E_k)~\delta(\Delta E_l) \no \\
&=& \delta_{kl}~\frac{1}{V^2}~(2\pi)^4~\delta^3(\Delta\vec{p})~\delta(\Delta E_k).
\end{eqnarray}
Substituting Eq.(3.25) into Eq.(3.5), one obtains
% (3.26)
\begin{eqnarray}
P_{\pi \to \mu\nu_{\mu} \to \mu\nu_{\alpha}} 
~=~ \frac{1}{V^2}~(2\pi)^4~\sum_k U_{\mu k}^2 U_{\alpha k}^2~
    \delta^3(\vec{p}_{\nu}+\vec{p}_{\mu}-\vec{p}_{\pi})~
    \delta(E_k+E_{\mu}-E_{\pi}), \no \\
\end{eqnarray}
or, in the two-generation case,
\begin{eqnarray*}
P_{\pi \to \mu\nu_{\mu} \to \mu\nu_{\mu}} 
&=& \frac{1}{V^2}~(2\pi)^4~\delta^3(\vec{p}_{\nu}+\vec{p}_{\mu}-\vec{p}_{\pi}) \\
&&  \{\sin^4\theta~\delta(E_1+E_{\mu}-E_{\pi})+\cos^4\theta~\delta(E_2+E_{\mu}-E_{\pi})\}, \\
P_{\pi \to \mu\nu_{\mu} \to \mu\nu_e} 
&=& \frac{1}{V^2}~(2\pi)^4~\delta^3(\vec{p}_{\nu}+\vec{p}_{\mu}-\vec{p}_{\pi}) \\
&&  \sin^2\theta\cos^2\theta~\{\delta(E_1+E_{\mu}-E_{\pi})+\delta(E_2+E_{\mu}-E_{\pi})\}.
\end{eqnarray*}
It is seen that energy-momentum conservation prevents different 
mass-eigenstates to interfere with one another and, as a result, 
each of mass-eigenstates appears to contribute separately to the 
transition probabilities.\footnote
{
An intuitive argument on connection between neutrino oscillation and 
energy-momentum conservation has been given by Kayser\cite{5} and 
by Lipkin\cite{18}.  
}

It is also interesting to examine the cases in which one of $\sigma_{\nu}$, 
$\sigma_{\mu}$ and $\sigma_{\pi}$ is kept finite. From Eq.(3.15), 
with $\sigma_{\pi}$ kept finite, one finds\footnote
{
$\vec{x}_{\kappa\lambda} \equiv \vec{x}_{\kappa}-\vec{x}_{\lambda}$, 
$t_{\kappa\lambda} \equiv t_{\kappa}-t_{\lambda}$, 
$\kappa,\lambda = \nu,\mu,\pi$. Note that, with 
$\sigma_{\nu},\sigma_{\mu} \to 0$, one has $\sigma \to \sigma_{\pi}$, 
$\Delta\tilde{E}_k \to \Delta E_k-\vec{v}_{\pi}\Delta\vec{p}$, 
$\bra\vec{X}\ket_k \to \vec{X}_{\pi}$ and $a_k,b_k,c_k \to 0$, 
while $t_k$ remains finite. 
}
\begin{eqnarray*}
\lim_{\sigma_{\mu},\sigma_{\nu} \To 0}~A_k &=& \frac{1}{V}~N_{\pi}~(\frac{2\pi}{\sigma_{\pi}^2})^{3/2}~(2\pi)\delta(\Delta E_k-\vec{v}_{\pi}\Delta\vec{p}) \no \\
&& \exp\{i(\vec{p}_{\nu}\vec{x}_{\nu\pi}-E_kt_{\nu\pi}+\vec{p}_{\mu}\vec{x}_{\mu\pi}-E_{\mu}t_{\mu\pi}) -\frac{1}{2\sigma_{\pi}^2}(\Delta\vec{p})^2\},
\end{eqnarray*}
from which one derives
%
% (3.27)
\begin{eqnarray}
\lim_{\sigma_{\mu},\sigma_{\nu} \To 0}~G_{kl} &=& \frac{1}{T}~\frac{1}{V^2}~N_{\pi}^2~(\frac{2\pi}{\sigma_{\pi}^2})^3~(2\pi)^2\delta(\Delta E_k-\vec{v}_{\pi}\Delta\vec{p})\delta(\Delta E_l-\vec{v}_{\pi}\Delta\vec{p}) \no \\
&& \exp\{-iE_{[kl]}t_{\nu\pi} -\frac{1}{\sigma_{\pi}^2}(\Delta\vec{p})^2\} \no \\
&=& \delta_{kl}~\frac{1}{V^2}~(2\pi)^4~(\frac{1}{\pi\sigma_{\pi}^2})^{3/2}~\delta(\Delta E_k-\vec{v}_{\pi}\Delta\vec{p})~\exp\{-\frac{1}{\sigma_{\pi}^2}(\Delta\vec{p})^2\}. \no \\
\end{eqnarray}
It is interesting to see that there appears in the amplitude 
a single $\delta$-function which implies that energies are conserved 
in the rest frame of the pion and this $\delta$-function eliminates 
interference terms from the transition probability. In contrast, 
with $\sigma_{\nu}$ kept finite, one has
% (3.28)
\begin{eqnarray}
\lim_{\sigma_{\pi},\sigma_{\mu} \To 0}~A_k &=& \frac{1}{V}~N_{\nu}~(\frac{2\pi}{\sigma_{\nu}^2})^{3/2}~(2\pi)\delta(\Delta E_k-\vec{v}_k\Delta\vec{p}) \no \\
&& \exp\{i(\vec{p}_{\mu}\vec{x}_{\mu\nu}-E_{\mu}t_{\mu\nu}-\vec{p}_{\pi}\vec{x}_{\pi\nu}+E_{\pi}t_{\pi\nu})-\frac{1}{2\sigma_{\nu}^2}(\Delta\vec{p})^2\}, \no \\
\lim_{\sigma_{\pi},\sigma_{\mu} \To 0}~G_{kl} &=& \frac{1}{V^2}~(2\pi)^4~(\frac{1}{\pi\sigma_{\nu}^2})^{3/2}~\frac{2\pi}{T}~\delta(\Delta E_k-\vec{v}_k\Delta\vec{p})\delta(\Delta E_l-\vec{v}_l\Delta\vec{p}) \no \\
&& \exp\{-\frac{1}{\sigma_{\nu}^2}(\Delta\vec{p})^2\}.
\end{eqnarray}
Here, remarkably, interference terms remain nonvanishing, but appear 
not to contain an oscillating factor.\footnote
{
$\Delta E_k-\vec{v}_k\Delta\vec{p}=0$ and 
$\Delta E_l-\vec{v}_l\Delta\vec{p}=0$ are not incompatible with each 
other and the product of the two $\delta$-functions, 
$\delta(\Delta E_k-\vec{v}_k\Delta\vec{p})
\delta(\Delta E_l-\vec{v}_l\Delta\vec{p})$, may be rewritten as 
$\delta(\Delta E_k-\vec{v}_k\Delta\vec{p})
\delta((E_k-E_l)(E_k+E_l+E_{\mu}-E_{\pi})/E_k)$. Note also that, if one 
lets $\sigma_{\pi} \To 0$ in Eq.(3.27) or lets $\sigma_{\nu} \To 0$ 
in Eq.(3.28), these equations reduce to Eq.(3.25).
}
\subsection*{3.4. Total transition probability}
From Eqs.(2.4) and (2.8), noting that the function 
$f(\vec{p};\vec{p}_{\nu},\sigma_{\nu})$, defined by Eq.(2.5), satisfies
\[
\int d^3p_{\nu} |f(\vec{p};\vec{p}_{\nu},\sigma_{\nu})|^2 ~=~ 1,
\]
one may derive such a relation as\footnote
{
$\vec{x}_{\lambda 0}' = \vec{x}_{\lambda}-\vec{x}_0',~t_{\lambda 0}' = t_{\lambda}-t_0',~\kappa = \nu,\mu,\pi$.
}
% (3.29)
\begin{eqnarray}
\lefteqn{\int d^3p_{\nu}\int d^3x_{\nu}~
\phi_k(\vec{x}_{\nu 0},t_{\nu 0};\vec{p}_{\nu},\sigma_{\nu})
\phi_l^*(\vec{x}_{\nu 0}',t_{\nu 0}';\vec{p}_{\nu},\sigma_{\nu})} \hspace{1cm} \no \\
&=& \int d^3p_{\nu} \int_Vd^3x_{\nu}~\lim_{\sigma_{\nu} \To 0}
[\phi_k(\vec{x}_{\nu 0},t_{\nu 0};\vec{p}_{\nu},\sigma_{\nu})
\phi_l^*(\vec{x}_{\nu 0}',t_{\nu 0}';\vec{p}_{\nu},\sigma_{\nu})].
\end{eqnarray}
On substituting Eq.(3.3), we integrate Eq.(3.7) with respect to 
$\vec{x}_{\nu}$ and $\vec{p}_{\nu}$ and consult Eq.(3.29), to obtain
% (3.30)
\begin{eqnarray}
\lefteqn{\int d^3p_{\nu} \int d^3x_{\nu}~
G_{kl}} \hspace{1cm} \no \\ 
&=& \frac{1}{T} \int d^3x_0 \int dt_0 \int d^3x_0' \int dt_0'~
    \phi_{\pi}^*(\vec{x}_{\pi 0},t_{\pi 0};\vec{p}_{\pi},\sigma_{\pi}) 
    \phi_{\pi}(\vec{x}_{\pi 0}',t_{\pi 0}';\vec{p}_{\pi},\sigma_{\pi}) \no \\
&& \qquad\qquad\qquad\qquad\qquad\qquad\quad \phi_{\mu}(\vec{x}_{\mu 0},t_{\mu 0};\vec{p}_{\mu},\sigma_{\mu}) \phi_{\mu}^*(\vec{x}_{\mu 0}',t_{\mu 0}';\vec{p}_{\mu},\sigma_{\mu}) \no \\
&& \qquad~\int d^3p_{\nu} \int_Vd^3x_{\nu}~\lim_{\sigma_{\nu} \To 0}
[\phi_k(\vec{x}_{\nu 0},t_{\nu 0};\vec{p}_{\nu},\sigma_{\nu})
\phi_l^*(\vec{x}_{\nu 0}',t_{\nu 0}';\vec{p}_{\nu},\sigma_{\nu})] \no \\
&=& \int d^3p_{\nu} \int_V d^3x_{\nu}~\lim_{\sigma_{\nu} \To 0} G_{kl}.
\end{eqnarray}
Similarly, we obtain
% (3.31)
\begin{eqnarray}
\int d^3p_{\mu} \int d^3x_{\mu}~
G_{kl} &=& \int d^3p_{\mu} \int_V d^3x_{\mu}~\lim_{\sigma_{\mu} \To 0} G_{kl},
\end{eqnarray}
%
% (3.32)
\begin{eqnarray}
\int d^3p_{\mu} \int d^3x_{\mu}~\int d^3p_{\nu} \int d^3x_{\nu}~G_{kl} 
&=& \int d^3p_{\mu} \int_V d^3x_{\mu} \int d^3p_{\nu} \int_V d^3x_{\nu}~\lim_{\sigma_{\mu},\sigma_{\nu} \To 0} G_{kl}. \no \\
\end{eqnarray}
Equation (3.32) implies that, as far as the integrated transition 
probability is concerned, there is no difference between treating decay 
products as plane-waves or as wave-packets.

Integrating Eq.(3.5) with respect to $\vec{x}_{\nu}$, $\vec{p}_{\nu}$, 
$\vec{x}_{\mu}$ and $\vec{p}_{\mu}$, summing it up with respect to 
$\alpha$ and consulting Eqs.(3.32) and (3.27), we obtain the total 
transition probability or total decay rate as
% (3.33)
\begin{eqnarray}
P_{\pi \to \mu \nu_{\mu}} &\equiv& \sum_{\alpha}~\int d^3p_{\mu} \int d^3x_{\mu}
\int d^3p_{\nu} \int d^3x_{\nu}~P_{\pi \to \mu\nu_{\mu} \to \mu\nu_{\alpha}} \no \\
&=& \sum_kU_{\mu k}^2~\int d^3p_{\mu} \int_V d^3x_{\mu}~\int d^3p_{\nu} \int_V d^3x_{\nu}~\lim_{\sigma_{\mu},\sigma_{\nu} \To 0}G_{kk} \no \\
&=& (2\pi)^4~(\frac{1}{\pi\sigma_{\pi}^2})^{3/2} \sum_k~U_{\mu k}^2~
    \int d^3p_{\mu} \int d^3p_{\nu} \no \\
&&  \delta(\Delta E_k-\vec{v}_{\pi}\Delta\vec{p})~\exp\{-\frac{1}{\sigma_{\pi}^2}
    (\Delta\vec{p})^2\},
\end{eqnarray}
which reduces, in the limit of $\sigma_{\pi} \to 0$, to
% (3.34)
\begin{eqnarray}
P_{\pi \to \mu\nu_{\mu}} &=& (2\pi)^4~\sum_k~U_{\mu k}^2~\int d^3p_{\mu} \int d^3p_{\nu} \no \\
&& \delta^3(\vec{p}_{\nu}+\vec{p}_{\mu}-\vec{p}_{\pi})~\delta(E_k+E_{\mu}-E_{\pi}).
\end{eqnarray}
Equation (3.34) as well as Equation (3.26) are expected results, 
which serves as a check of consistency of our treatment as a whole.
\section{An approximate treatment}
In this section, we like to propose an approximate prescription to 
handle with the trajectory factor $\exp\{-\sigma^2Z_{kl}\}$, 
the energy-momentum factor $\exp\{-H_{kl}/\sigma^2\}$ and 
the phase factor $\exp\{-i\Theta_{kl}\}$.\footnote
{
We shall introduce quantities with suffix $(kl)$ and quantities with 
a prime or a double prime; the former are, by definition, quantities 
depending on $\vec{v}_{(kl)} \equiv (\vec{v}_k+\vec{v}_l)/2$ and/or 
$E_{(kl)} \equiv (E_k+E_l)/2$, just as the quantities with suffix $k$ 
depend on $\vec{v}_k$ and/or $E_k$, while the latter are defined 
explicitly in Appendix B and vanish if $k \neq l$ or in the limit of 
$\sigma_{\nu} \to 0$. Also, many of equations given in this section 
will be derived explicitly in Appendix B.
}
\subsection*{4.1. Trajectory factor}
As noted before, the factor $\exp\{-\sigma^2Z_{kk}\} 
\equiv \exp\{-\sigma^2(c_k-a_kt_k^2)\}$, contained in $G_{kk}$, 
may be interpreted as something which reflects that each of the 
particles involved propagates nearly along its classical trajectory, 
Eq.(3.21), while, for $k \neq l$, $c_k-a_kt_k^2 = 0$ and 
$c_l-a_lt_l^2 = 0$ are not compatible with each other and 
$Z_{kl} \equiv (c_k-a_kt_k^2+c_l-a_lt_l^2)/2 = 0$ never holds. 
We have encountered a similar situation in Sec.2. There, writing
% (4.1)
\begin{eqnarray}
\vec{v}_{k,l} ~=~ \vec{v}_{(kl)} \pm \frac{1}{2}\vec{v}_{[kl]},
\end{eqnarray}
and, accordingly,
\begin{eqnarray*}
\frac{1}{2}((\vec{x}_{\nu 0}-\vec{v}_kt_{\nu 0})^2+(\vec{x}_{\nu 0}-\vec{v}_lt_{\nu 0})^2)~=~ (\vec{x}_{\nu 0}-\vec{v}_{(kl)}t_{\nu 0})^2+\frac{1}{4}(\vec{v}_{[kl]})^2t_{\nu 0}^2,
\end{eqnarray*}
we have reasonably regarded Eq.(2.12) as a classical trajectory of 
the interfering neutrinos. Here, on writing $Z_{kl}$ as
% (4.2)
\begin{eqnarray}
Z_{kl} ~=~ c_{(kl)}-a_{(kl)}t_{(kl)}^2 + \zeta_{kl},
\end{eqnarray}
and, accordingly, $\exp\{-\sigma^2Z_{kl}\}$ as
% (4.3)
\begin{eqnarray}
\exp\{-\sigma^2Z_{kl}\} ~=~ z_{kl}~\xi_{kl}^{(1)},
\end{eqnarray}
where
% (4.4)
\begin{eqnarray}
z_{kl} &=& \exp\{-\sigma^2(c_{(kl)}-a_{(kl)}t_{(kl)}^2)\}, \no \\
\xi_{kl}^{(1)} &=& \exp\{-\sigma^2\zeta_{kl}\},
\end{eqnarray}
we shall regard 
% (4.5)
\begin{eqnarray}
c_{(kl)}-a_{(kl)}t_{(kl)}^2 ~=~ 0
\end{eqnarray}
as something which reflecs that each of the particles involved, 
including the interfering neutrinos, propagates along its classical 
trajectory and hence $z_{kl}$ as a factor representing that each of 
the particles involved propagates nearly along its classical trajectory.

$\zeta_{kl}$, defined by Eq.(4.2), is given explicitly by
% (4.6)
\begin{eqnarray}
\zeta_{kl} &\equiv& \frac{1}{2}(c_k-a_kt_k^2+c_l-a_lt_l^2) - (c_{(kl)}-a_{(kl)}t_{(kl)}^2) \no \\
&=& \frac{1}{a_{(kl)}((a_{(kl)}+a_{kl}'')^2-(a_{kl}')^2)} \no \\
&& \times [~(a_{(kl)}+a_{kl}'')\{a_{(kl)}^2c_{kl}''-2a_{(kl)}b_{(kl)}b_{kl}''+b_{(kl)}^2a_{kl}''+a_{(kl)}(a_{kl}''c_{kl}''-(b_{kl}'')^2)\} \nonumber \\
&& ~-(a_{(kl)}b_{kl}'-b_{(kl)}a_{kl}')^2
-a_{(kl)}(c_{kl}''(a_{kl}')^2-2b_{kl}''a_{kl}'b_{kl}'+a_{kl}''(b_{kl}')^2)~].
\end{eqnarray}
When Equation (4.5) holds, which implies
% (4.7)
\begin{eqnarray}
\vec{X}_{[\kappa\lambda]} ~=~ -\vec{v}_{[\kappa\lambda]}\overline{t}_0, \qquad 
\kappa,~\lambda ~=~ (kl),~\mu,~\pi,
\end{eqnarray}
$\overline{t}_0$ being a constant independent of $\kappa$ and 
$\lambda$, $\zeta_{kl}$ reduces to 
$\overline{\zeta}_{kl}(t_{\nu}-\overline{t}_0)^2$, where
% (4.8)
\begin{eqnarray}
\overline{\zeta}_{kl} &=& \frac{(a_{(kl)}+a_{kl}'')(a_{(kl)}a_{kl}''-\frac{1}{4}(a_{kl}')^2)}{(a_{(kl)}+a_{kl}'')^2-(a_{kl}')^2},
\end{eqnarray}
and the trajectory factor $\exp\{-\sigma^2Z_{kl}\}$ may be approximated 
as
% (4.9)
\begin{eqnarray}
\exp\{-\sigma^2Z_{kl}\} ~\approx~ z_{kl}~\overline{\xi}_{kl}^{(1)},
\end{eqnarray}
where
% (4.10)
\begin{eqnarray}
\overline{\xi}_{kl}^{(1)} ~=~ \exp\{-\sigma^2\overline{\zeta}_{kl}(t_{\nu}-\overline{t}_0)^2\}.
\end{eqnarray}

Necessary conditions for interference terms to be appreciable follow 
from Eqs.(4.3), (4.4), (4.9) and (4.10):
\begin{enumerate}
\item[(a)]
$c_{(kl)}-a_{(kl)}t_{(kl)} ~\lesssim~ 1/\sigma^2$; and
\item[(b)]
$\zeta_{kl}~~{\rm or}~~\overline{\zeta}_{kl}(t_{\nu}-\overline{t}_0)^2 ~\lesssim~ 1/\sigma^2$.
\end{enumerate}
The former condition applies to the diagonal terms $G_{kk}$ too and 
implies that each of the particles involved should propagate along its 
classical trajectory at least approximately, or that those 
mass-eigenstates which do not satisfy this condition contribute little 
to the transition probability, while the latter is a condition which 
corresponds to Eq.(2.15), known as coherent condition for neutrino 
oscillation\cite{15}, and implies that those two mass-eigenstates, 
if their masses  do not satisfy this condition, interfere little 
with each other. 

From Eq.(4.7), it follows that
% (4.11)
\begin{eqnarray}
\vec{x}_{\nu}-\vec{v}_{(kl)}(t_{\nu}-\overline{t}_0) ~=~ 
\vec{x}_{\mu}-\vec{v}_{\mu}(t_{\mu}-\overline{t}_0) ~=~ 
\vec{x}_{\pi}+\vec{v}_{\pi}(\overline{t}_0-t_{\pi}) ~\equiv~ \vec{\overline{x}}_0.
\end{eqnarray}
Since these equations appear to be exactly similar to Eq.(3.21) 
which involves the decay space-time point $(\vec{x}_0,t_0)$, 
we shall refer to $(\overline{\vec{x}}_0,\overline{t}_0)$ as 
pseudo decay space-time point. Note that the pseudo decay time 
$\overline{t}_0$, and hence the pseudo traveling time of the neutrinos 
$t_{\nu}-\overline{t}_0$ as well, may in principle be deduced from 
knowledge of ($\vec{x}_{\nu}, t_{\nu}$) and $\vec{v}_{(kl)}$ and of 
($\vec{x}_{\mu}, t_{\mu}$) and $\vec{v}_{\mu}$ (or of 
($\vec{x}_{\pi}, t_{\pi}$) and $\vec{v}_{\pi}$).
\subsection*{4.2. Energy-momentum factor}
In the limit of $\sigma \to 0$, $\exp\{-H_{kk}/\sigma^2\}$ 
gives rise to $\delta(E_k+E_{\mu}-E_{\pi})
\delta^3(\vec{p}_{\nu}+\vec{p}_{\mu}-\vec{p}_{\pi})$, 
while $\exp\{-H_{kl}/\sigma^2\}$ with $k \neq l$ vanishes 
(see also Sec.3.3).\footnote
{
More precisely, one has
\begin{eqnarray*}
\lim _{\sigma \to 0}~\exp\{-\frac{1}{\sigma^2}H_{kl}\} &=& 
\delta_{kl}~(\pi\sigma^2)^{3/2}~(\pi\sigma^2a_k)^{1/2}~\delta^3(\Delta\vec{p})~\delta(\Delta E_k),
\end{eqnarray*}
and, accordingly,
\begin{eqnarray*}
\lim_{\sigma_{\pi},\sigma_{\mu},\sigma_{\nu} \To 0}~G_{kl} &=& \delta_{kl}~\frac{1}{V^2}~\frac{1}{T}~(\frac{\pi}{\sigma^2a_k})^{1/2}~(2\pi)^4~\delta^3(\Delta\vec{p})~\delta(\Delta E_k).
\end{eqnarray*}
In order for this expression to coincide with Eq.(3.25), one needs to 
identify $T$ with $(\pi/\sigma^2a_k)^{1/2}$.
}
 In order for interference terms to remain nonvanishing and appreciable, 
the two constituting factors of $\exp\{-H_{kl}/\sigma^2\}$, 
$\exp\{-(\Delta\vec{p})^2/2\sigma^2-(\Delta\tilde{E}_k)^2/2\sigma^2a_k\}$ 
and $\exp\{-(\Delta\vec{p})^2/2\sigma^2 - (\Delta\tilde{E}_l)^2/2\sigma^2a_l\}$, have to overlap appreciably with each other. 

To look into this condition more deeply, we rewrite 
$(\Delta\tilde{E}_k)^2/2a_k+(\Delta\tilde{E}_l)^2/2a_l$ as
% (4.12)
\begin{eqnarray}
\frac{1}{2a_k}(\Delta\tilde{E}_k)^2+\frac{1}{2a_l}(\Delta\tilde{E}_l)^2 
&=& \frac{1}{a_{(kl)}}(\Delta\tilde{E}_{(kl)})^2 + \eta_{kl},
\end{eqnarray}
and, accordingly, $\exp\{-H_{kl}/\sigma^2\}$ as
% (4.13)
\begin{eqnarray}
\exp\{-\frac{1}{\sigma^2}H_{kl}\} &=& h_{kl}~\xi_{kl}^{(2)},
\end{eqnarray}
where
% (4.14)
\begin{eqnarray}
h_{kl} &=& \exp\{-\frac{1}{\sigma^2}((\Delta\vec{p})^2 + \frac{1}{a_{(kl)}}(\Delta\tilde{E}_{(kl)})^2)\}, \no \\
\xi_{kl}^{(2)} &=& \exp\{-\frac{1}{\sigma^2}\eta_{kl}\}.
\end{eqnarray}
The factor $h_{kl}$ has a peak at
% (4.15)
\begin{eqnarray}
\Delta\vec{p} &\equiv& \vec{p}_{\nu}+\vec{p}_{\mu}-\vec{p}_{\pi} ~=~ 0, ~\no \\
\Delta E_{(kl)} &\equiv& E_{(kl)}+E_{\mu}-E_{\pi} ~=~ 0,
\end{eqnarray}
which may reasonably be regarded as representing the energy-momentum 
conservation in the presence of neutrino mixing. 

$\eta_{kl}$, defined by Eq.(4.12), is given explicitly by
% (4.16)
\begin{eqnarray}
\eta_{kl} &\equiv& \frac{1}{2a_k}(\Delta\tilde{E}_k)^2+\frac{1}{2a_l}(\Delta\tilde{E}_l)^2 - \frac{1}{a_{(kl)}}(\Delta\tilde{E}_{(kl)})^2 \no \\
&=& \frac{1}{4a_{(kl)}((a_{(kl)}+a_{kl}'')^2-(a_{kl}')^2)} \no \\
&& \times [~(a_{(kl)}(a_{(kl)}+a_{kl}'')(\Delta\tilde{E}_{kl}')^2-4a_{(kl)}a_{kl}'\Delta\tilde{E}_{(kl)}\Delta\tilde{E}_{kl}' \no \\
&& ~+4((a_{kl}')^2-a_{kl}''(a_{(kl)}+a_{kl}''))(\Delta\tilde{E}_{(kl)})^2~].
\end{eqnarray}
When Equation (4.15) holds, $\eta_{kl}$ reduces to
% (4.17)
\begin{eqnarray}
\hat{\eta}_{kl} &=& \frac{a_{(kl)}+a_{kl}''}{4((a_{(kl)}+a_{kl}'')^2-(a_{kl}')^2)}(E_{[kl]})^2,
\end{eqnarray}
and $\exp\{-H_{kl}/\sigma^2\}$ may be approximated as
%
% (4.18)
\begin{eqnarray}
\exp\{-\frac{1}{\sigma^2}H_{kl}\} &\approx& h_{kl}~\hat{\xi}_{kl}^{(2)},
\end{eqnarray}
where
% (4.19)
\begin{eqnarray}
\hat{\xi}_{kl}^{(2)} &=& \exp\{-\frac{1}{\sigma^2}\hat{\eta}_{kl}\}.
\end{eqnarray}
Necessary conditions for interference terms to be appreciable now 
follow from Eqs.(4.13), (4.14), (4.18) and (4.19):
\begin{itemize}
\item[(c)]
$(\Delta\vec{p})^2 + (\Delta\tilde{E}_{(kl)})^2/a_{(kl)} ~\lesssim~ \sigma^2$; and 
\item[(d)]
$\eta_{kl}$ ~or~ $\hat{\eta}_{kl} ~\lesssim~ \sigma^2$.
\end{itemize}
The former condition applies to the diagonal terms $G_{kk}$ too and 
implies that energies and momenta involved should conserved at least 
approximately, or that those mass-eigenstates which do not satisfy 
this condition contribute little to the transition probability, 
while the latter condition implies that those two mass-eigenstates, 
if their masses  do not satisfy this condition, interfere little with 
each other. It is to be noted also that $\xi_{kl}^{(2)}$ or 
$\hat{\xi}_{kl}^{(2)}$ seems to have something to do with the 
suppression factor, given by 
$\exp\{-(E_{[kl]})^2/2\sigma_{\nu}^2(\vec{v}_k^2+\vec{v}_l^2)\}$, 
which appears in a more general wave-packet treatment of neutrino 
oscillation (see Appendix A).
\subsection*{4.3. Phase factor}
$\Theta_{kl}$, Eq.(3.20), may be rewritten as
% (4.20)
\begin{eqnarray}
\Theta_{kl} &=& \Delta\tilde{E}_{kl}'(t_{\nu}-\frac{1}{2}(t_k+t_l)) - \Delta\tilde{E}_{(kl)}(t_k-t_l) \nonumber \\
&=& \Delta\tilde{E}_{kl}'\{t_{\nu}+\frac{(a_{(kl)}+a_{kl}'')(b_{(kl)}+b_{kl}'')-a_{kl}'b_{kl}'}{(a_{(kl)}+a_{kl}'')^2-(a_{kl}')^2}\} \nonumber \\
&& + 2\Delta\tilde{E}_{(kl)}\frac{(b_{kl}'(a_{(kl)}+a_{kl}'')-a_{kl}'(b_{(kl)}+b_{kl}'')}{(a_{(kl)}+a_{kl}'')^2-(a_{kl}')^2}.
\end{eqnarray}
When Equation (4.7) holds, $\Theta_{kl}$ reduces to 
$\overline{\Theta}_{kl}(t_{\nu}-\overline{t}_0)$, where
% (4.21)
\begin{eqnarray}
\overline{\Theta}_{kl} &=& \Delta\tilde{E}_{kl}' \frac{a_{(kl)}(a_{(kl)}+a_{kl}'')-(a_{kl}')^2/2}{(a_{(kl)}+a_{kl}'')^2-(a_{kl}')^2} -\Delta\tilde{E}_{(kl)} \frac{a_{kl}'(a_{(kl)}-a_{kl}'')}{(a_{(kl)}+a_{kl}'')^2-(a_{kl}')^2}. \no \\
\end{eqnarray}
If Equations (4.15) are further applied, $\Theta_{kl}$ reduces to 
$\hat{\overline{\Theta}}_{kl}(t_{\nu}-\overline{t}_0)$, where
% (4.22)
\begin{eqnarray}
\hat{\overline{\Theta}}_{kl} &=& \frac{a_{(kl)}(a_{(kl)}+a_{kl}'')-(a_{kl}')^2/2}{(a_{(kl)}+a_{kl}'')^2-(a_{kl}')^2} E_{[kl]}.
\end{eqnarray}
\subsection*{4.4. Summary}
We have rewritten Eq.(3.16),
\[
G_{kl} ~\equiv~ N_{kl}~\exp\{-\sigma^2Z_{kl}-\frac{1}{\sigma^2}H_{kl}-i\Theta_{kl}\},
\]
as
% (4.23)
\begin{eqnarray}
G_{kl} &=& N_{kl}~z_{kl}~h_{kl}~\xi_{kl}^{(1)}~\xi_{kl}^{(2)}~\exp\{-i\Theta_{kl}\},
\end{eqnarray}
where $z_{kl}$, $h_{kl}$, $\xi_{kl}^{(1)}$ and $\xi_{kl}^{(2)}$ are 
given by Eqs.(4.4) and (4.14). Substituting Eqs.(3.16) and (4.23) into 
Eq.(3.5), one obtains
% (4.24)
\begin{eqnarray}
P_{\pi \to \mu\nu_{\mu} \to \mu\nu_{\alpha}} 
&=& \sum_{k,l}U_{\mu k}U_{\alpha k}U_{\mu l}U_{\alpha l}~N_{kl}~
\exp\{-\sigma^2Z_{kl}-\frac{1}{\sigma^2}H_{kl}-i\Theta_{kl}\} \no \\
&=& \sum_{k,l}U_{\mu k}U_{\alpha k}U_{\mu l}U_{\alpha l}~N_{kl}~
z_{kl}~h_{kl}~\xi_{kl}^{(1)}~\xi_{kl}^{(2)} \cos(\Theta_{kl}),
\end{eqnarray}
We have then shown that Equation (4.24) may be approximated as 
% (4.25)
\begin{eqnarray}
P_{\pi \to \mu\nu_{\mu} \to \mu\nu_{\alpha}} 
&\approx& \sum_{k,l}U_{\mu k}U_{\alpha k}U_{\mu l}U_{\alpha l}~N_{kl}~z_{kl}~h_{kl}~\overline{\xi}_{kl}^{(1)}~\hat{\xi}_{kl}^{(2)} \cos(\hat{\overline{\Theta}}_{kl}(t_{\nu}-\overline{t}_0)), \no \\
\end{eqnarray}
where $\overline{\xi}_{kl}^{(1)}$, $\hat{\xi}_{kl}^{(2)}$, 
and $\hat{\overline{\Theta}}_{kl}$ are given respectively by 
Eqs.(4.10), (4.19) and (4.22). 

We have furthermore discussed implications of each factor 
in Eqs.(4.24) and (4.25) and derived necessary conditions 
for oscillating terms to be appreciable. It is to be emphasized here 
that the two features implied respectively by $z_{kl}$ and $h_{kl}$, 
that is, each of the particles involved propagates nearly along its 
classical trajectory on the one hand and energies and momenta 
of the particles involved are nearly conserved on the other hand, 
are complementary to each other in the sense that, as $\sigma$ becomes 
larger (smaller), the former (latter) becomes more prominent, 
and that the two suppression factors $\xi_{kl}^{(1)}$ or 
$\overline{\xi}_{kl}^{(1)}$ and $\xi_{kl}^{(2)}$ or $\hat{\xi}^{(2)}$ 
are also complementary to each other in the sense that, as $\sigma^2$ 
becomes larger (smaller), the former (latter) becomes more effective.

If $c_{(kl)}-a_{(kl)}t_{(kl)}^2 \ll 1/\sigma^2$ is assumed to hold 
for any $k$ and $l$, Equations (4.24) and (4.25) may be approximated as
% (4.26)
\begin{eqnarray}
P_{\pi \to \mu\nu_{\mu} \to \mu\nu_{\alpha}} &\simeq& 
\sum_{k,l}U_{\mu k}U_{\alpha k}U_{\mu l}U_{\alpha l}~N_{kl}~h_{kl}~\overline{\xi}_{kl}^{(1)}~\xi_{kl}^{(2)} \cos(\overline{\Theta}_{kl}(t_{\nu}-\overline{t}_0)) \no \\
&\approx& \sum_{k,l}U_{\mu k}U_{\alpha k}U_{\mu l}U_{\alpha l}~N_{kl}~h_{kl}~\overline{\xi}_{kl}^{(1)}~\hat{\xi}_{kl}^{(2)} \cos(\hat{\overline{\Theta}}_{kl}(t_{\nu}-\overline{t}_0)),
\end{eqnarray}
where $\overline{\Theta}_{kl}$ is given by (4.21). Equations (4.25) and 
(4.26) are to be contrasted with Eqs.(2.10) and (2.13) and also with 
Eq.(3.26).
\section{Comparison with the approach by Dolgov et al.}
In the wave-packet treatment of $\pi \to \mu\nu$ decay given in 
\cite{13}, assuming that a pion has been created with some momentum 
distribution $f(\vec{p}_{\pi})$, the authors define a transition 
amplitude and the corresponding transition probability by
\begin{eqnarray*}
A_{\pi \to \mu\nu_{\mu} \to \mu\nu_{\alpha}}^D(\vec{x}_0,t_0) 
~=~ \sum_k U_{\mu k}U_{\alpha k} \int d^3p_{\pi}~f(\vec{p}_{\pi})~\exp\{i\varphi_k^D(\vec{p}_{\pi})\},
\end{eqnarray*}
and
% (5.1)
\begin{eqnarray}
P_{\pi \to \mu\nu_{\mu} \to \mu\nu_{\alpha}}^D(\vec{x}_0,t_0) 
&=& |A_{\pi \to \mu\nu_{\mu} \to \mu\nu_{\alpha}}^D(\vec{x}_0,t_0)|^2 \no \\
&=& \sum_{k,l} U_{\mu k}U_{\alpha k}U_{\mu l}U_{\alpha l}~G_{kl}^D(\vec{x}_0,t_0),
\end{eqnarray}
where
\begin{eqnarray*}
G_{kl}^D(\vec{x}_0,t_0) 
~=~ \int d^3p_{\pi} \int d^3p_{\pi}' f(\vec{p}_{\pi}) f(\vec{p}_{\pi}') \exp\{i(\varphi_k^D(\vec{p}_{\pi})-\varphi_l^D(\vec{p}_{\pi}'))\},
\end{eqnarray*}
with
\begin{eqnarray*}
\varphi_k^D(\vec{p}_{\pi}) &=& \vec{p}_{\nu k}\vec{x}_{\nu 0}-E_k(\vec{p}_{\nu k})t_{\nu 0}+\vec{p}_{\mu k}\vec{x}_{\mu 0}-E_{\mu}(\vec{p}_{\mu k})t_{\mu 0}-\vec{p}_{\pi}\vec{x}_{\pi 0}+E_{\pi}(\vec{p}_{\pi})t_{\pi 0}, \\
\varphi_l^D(\vec{p}_{\pi}') &=& \vec{p}_{\nu l}'\vec{x}_{\nu 0}-E_l(\vec{p}_{\nu l}')t_{\nu 0}+\vec{p}_{\mu l}'\vec{x}_{\mu 0}-E_{\mu}(\vec{p}_{\mu l}')t_{\mu 0}-\vec{p}_{\pi}'\vec{x}_{\pi 0}+E_{\pi}(\vec{p}_{\pi}')t_{\pi 0}.
\end{eqnarray*}
Here, the energy and momentum of the neutrino mass-eigenstate and of 
the muon are all regarded as depending, through the energy-momentum 
conservation law, on the mass of the neutrino and hence carrying the 
suffix $k$ or $l$. Assuming that the dispersion of $f(\vec{p}_{\pi})$ 
is small and that each of the particles involved should propagate 
along its classical trajectory (cf. Eq.(3.21)), they claim that 
Equation (5.1) reduces to
% (5.2)
\begin{eqnarray}
P_{\pi \to \mu\nu_{\mu} \to \mu\nu_{\alpha}}^D(\vec{x}_0,t_0) 
&=& \sum_{k,l} U_{\mu k}U_{\alpha k}U_{\mu l}U_{\alpha l}~\cos(\frac{m_k^2-m_l^2}{2E_{\nu}}t_{\nu 0}).
\end{eqnarray}

Although we may also define an $(\vec{x}_0,t_0)$-dependent transition 
probability from Eqs.(3.2) and (3.3), we like to point out here that, 
in spite that the approach we have developed in Sec.3 and Sec.4 is 
characterized among other things by the space-time point of decay, 
$(\vec{x}_0,t_0)$, being integrated out, the transition probability 
we have derived, Eq.(4.26), seems to describe a situation very much 
close to what Dolgov et al. claim to describe with Eq.(5.2): 
Equation (4.26) is  derived with each of the particles involved 
assumed to propagate along its classical trajectory, and, as a result, 
becomes depending on the pseudo traveling time $t_{\nu}-\overline{t}_0$, 
which just corresponds to the traveling time $t_{\nu}-t_0$ 
in the $(\vec{x}_0,t_0)$-dependent probability (5.2) of Dolgov et al.. 
It is to be noted at the same time, however, that Equation (4.26) 
appears to be  distinct from Eq.(5.2) in the following three points: 
(a) each term in Eq.(4.26) contains a factor which implies that 
the energy-momentum conservation holds only approximately; 
(b) each of the oscillating factors in Eq.(4.26) is multiplied 
by the two suppression factors $\overline{\xi}_{kl}^{(1)}$ and 
$\hat{\xi}_{kl}^{(2)}$ which are absent in Eq.(5.2); and 
(c) the oscillation period in Eq.(4.26) deviates in general from the 
oscillation period $4\pi E_{\nu}/|m_k^2-m_l^2|$ in Eq.(5.2).

In addition to the case in which both neutrinos and muons are detected 
(Case A), Dolgov et al.\cite{13} consider also the case in which only 
muons are detected (Case B) and the case in which only neutrinos are 
detected (Case C). In our approach, the transition probability 
corresponding to Case B may be obtained by integrating Eq.(3.5) with 
respect to $\vec{x}_{\nu}$ and $\vec{p}_{\nu}$ and summing it up with 
respect to $\alpha$:
% (5.5)
\begin{eqnarray}
P_{\pi \to \mu \nu_{\mu} \to \mu} 
&\equiv& \sum_{\alpha}~\int d^3p_{\nu} \int d^3x_{\nu}~P_{\pi \to \mu\nu_{\mu} \to \mu\nu_{\alpha}} \no \\
&=& \sum_{\alpha,k,l} U_{\mu k}U_{\alpha k}U_{\mu l}U_{\alpha l}~\int d^3p_{\nu} \int d^3x_{\nu}~G_{kl} \no \\
&=& \sum_k U_{\mu k}^2~\int d^3p_{\nu} \int d^3x_{\nu}~G_{kk}.
\end{eqnarray}
As seen, interference terms drop out, implying that muons do not 
oscillate. Although our approach gives, not only for Case A but also 
for Case B and Case C, results more or less different from what Dolgov 
et al. claim, we share with them the conclusion that muons do not 
oscillate.\footnote
{
Note however that there have been some debates as regards whether 
or not muons can oscillate\cite{19}.
}
\section{Conclusions}
Our wave-packet approach to pion decay is characterized by treating 
all particles involved as wave-packets and by integrating out 
the space-time point of decay $(\vec{x}_0,t_0)$. We have seen that, 
in such an approach, (1) energy-momentum conservation 
appears to hold only approximately, (2) exact energy-momentum 
conservation, which holds in the plane-wave limit, would eliminate 
neutrino osillating terms from the transition probability, and (3) 
treating only one of the particles involved as a wave-packet is 
insufficient to allow for neutrino oscillating terms to appear. 
We have furthermore developed an approximate treatment, 
which allows one to define a pseudo decay space-time point 
$(\overline{\vec{x}}_0,\overline{t}_0)$ and to express the 
transition probability, originally given by Eq.(4.24), approximately 
as Eqs.(4.25) and (4.26).

To conclude, we like to mention that phenomenological implications of 
the outcomes of the present study need to be examined separately 
and that our approach may readily be applied to various decay as well 
as production processes. \\ \\
{\large \bf Acknowledgments} \\
The authors are indebted to Professor S.Kamefuchi, Professor M.Obu and 
the members of particle physics group at Nihon University for 
discussions and encouragement.
%
%%% Appendix %%%
%
\section*{Appendix}
\setcounter{equation}{0}
\def\theequation{A.\arabic{equation}}
\subsection*{Appendix A. A more general three-dimensional wave-packet 
treatment of neutrino oscillation}
More generally, one may conceive a situation in which the peak momentum 
of the neutrino mass-eigenstates, $\vec{p}_{\nu}$ in Eqs.(2.4) and 
(2.5), depends on the suffix $k$ \cite{11}, \cite{20}. If 
$\vec{p}_{\nu}$ in Eqs.(2.2) and (2.3) is replaced by $\vec{p}_k$, 
one would be led to
% (A.1)
\begin{eqnarray}
P_{\nu_{\mu} \rightarrow \nu_{\alpha}}(\vec{x}_{\nu 0},t_{\nu 0}) 
&=& N_{\nu}^2~\sum_{k,l} U_{\mu k} U_{\alpha k} U_{\mu l} U_{\alpha l}
~\exp\{i(\vec{p}_{[kl]}\vec{x}_{\nu 0}-E_{[kl]}t_{\nu 0})\} \no \\
&& \exp\{-\frac{1}{2}\sigma_{\nu}^2(\vec{x}_{\nu 0}-\vec{v}_kt_{\nu 0})^2-\frac{1}{2}\sigma_{\nu}^2(\vec{x}_{\nu 0}-\vec{v}_lt_{\nu 0})^2\} \no \\
&=& N_{\nu}^2~\sum_{k,l} U_{\mu k} U_{\alpha k} U_{\mu l} U_{\alpha l}~
\cos(\vec{p}_{[kl]}\vec{x}_{\nu 0}-E_{[kl]}t_{\nu 0}) \no \\
&& \exp\{-\sigma_{\nu}^2(\vec{x}_{\nu 0}-\vec{v}_{(kl)}t_{\nu 0})^2\}
\exp\{-\frac{1}{4}\sigma_{\nu}^2(\vec{v}_{[kl]})^2t_{\nu 0}^2\},
\end{eqnarray}
which may be approximated as
% (A.2)
\begin{eqnarray}
P_{\nu_{\mu} \rightarrow \nu_{\alpha}}(\vec{x}_{\nu 0},t_{\nu 0}) 
&\simeq& N_{\nu}^2~\sum_{k,l} U_{\mu k} U_{\alpha k} U_{\mu l} U_{\alpha l} \no \\
&& \cos\{(\vec{p}_{[kl]}\vec{v}_{(kl)}-E_{[kl]})t_{\nu 0}\}~
\exp\{-\frac{1}{4}\sigma_{\nu}^2(\vec{v}_{[kl]})^2t_{\nu 0}^2\},
\end{eqnarray}
where $\vec{p}_{[kl]} ~=~ \vec{p}_k-\vec{p}_l$. 
With a little algebra, one may verify that
% (A.3)
\begin{eqnarray}
\vec{p}_{[kl]}\vec{v}_{(kl)}-E_{[kl]} ~=~ -\frac{m_k^2-m_l^2}{2E_{(kl)}}
-\frac{\vec{p}_{[kl]}\vec{v}_{[kl]}}{4E_{(kl)}}E_{[kl]}.
\end{eqnarray}

In practical experiments, $D \equiv |\vec{x}_{\nu 0}|$ and $t_{\nu 0}$ 
are not measured in an independent way; instead, the former is measured,
 while the latter is inferred from the former. Such a situation 
corresponds to $t_{\nu}$ or $t_0$ to be integrated out in Eq.(A.1). 
One then obtains
% (A.4)
\begin{eqnarray}
\lefteqn{P_{\nu_{\mu} \to \nu_{\alpha}}(\vec{x}_{\nu 0}) 
~\equiv~ \int dt_{\nu}~P_{\nu_{\mu} \to \nu_{\alpha}}(\vec{x}_{\nu 0},t_{\nu 0})} \hspace{1cm} \no \\
&=& N_{\nu}^2~\sum_{k,l} U_{\mu k} U_{\alpha k} U_{\mu l} U_{\alpha l}~
(\frac{2\pi}{\sigma_{\nu}^2(\vec{v}_k^2+\vec{v}_l^2)})^{1/2}~\cos\{(\vec{p}_{[kl]}-\frac{2E_{[kl]}}{\vec{v}_k^2+\vec{v}_l^2}\vec{v}_{(kl)})\vec{x}_{\nu 0}\} \no \\
&& \exp\{-\sigma_{\nu}^2(\vec{x}_{\nu 0}^2-\frac{2(\vec{v}_{(kl)}\vec{x}_{\nu 0})^2}{\vec{v}_k^2+\vec{v}_l^2})\}~\exp\{-\frac{(E_{[kl]})^2}{2\sigma_{\nu}^2(\vec{v}_k^2+\vec{v}_l^2)}\}.
\end{eqnarray}
Equations (A.2) $\sim$ (A.4) coincide in the one-dimensional case with 
the corresponding equations derived in \cite{11} and \cite{20}).\footnote
{
In the case in which $\vec{p}_k$ and $\vec{x}_{\nu 0}$ may be expressed 
as $\vec{p}_k=\vec{p}_{\nu}|\vec{p}_k|/|\vec{p}_{\nu}|$ (for any $k$) 
and as $\vec{x}_{\nu 0}=\vec{p}_{\nu}|\vec{x}_{\nu 0}|/|\vec{p}_{\nu}|$,
 one has
\begin{eqnarray*}
\vec{p}_{[kl]}\vec{v}_{(kl)}-E_{[kl]} &=& -\frac{m_k^2-m_l^2}{2E_{(kl)}}
-\frac{(|\vec{p}_k|-|\vec{p}_l|)(|\vec{v}_k|-|\vec{v}_l|)}{4E_{(kl)}}E_{[kl]}, \\
\vec{p}_{[kl]}-\frac{2\vec{v}_{(kl)}}{\vec{v}_k^2+\vec{v}_l^2}E_{[kl]} 
&=& |\vec{p}_k|-|\vec{p}_l|-\frac{|\vec{v}_k|+|\vec{v}_l|}{|\vec{v}_k|^2+|\vec{v}_l|^2}E_{[kl]} \\
&=& -\frac{m_k^2-m_l^2}{|\vec{p}_k|+|\vec{p}_l|}-\frac{(|\vec{v}_k|-|\vec{v}_l|)(E_k|\vec{v}_l|-E_l|\vec{v}_k|)}{(|\vec{p}_k|+|\vec{p}_l|)(|\vec{v}_k|^2+|\vec{v}_l|^2)}E_{[kl]}, \\
\vec{x}_{\nu 0}^2-\frac{2(\vec{v}_{(kl)}\vec{x}_{\nu 0})^2}{\vec{v}_k^2+\vec{v}_l^2} 
&=& |\vec{x}_{\nu 0}|^2-\frac{((|\vec{v}_k|+|\vec{v}_l|)|\vec{x}_{\nu 0}|)^2}{2(|\vec{v}_k|^2+|\vec{v}_l|^2)} ~=~ \frac{(|\vec{v}_k|-|\vec{v}_l|)^2}{2(|\vec{v}_k|^2+|\vec{v}_l|^2)}|\vec{x}_{\nu 0}|^2.
\end{eqnarray*}
}
 It is interesting to observe the following.
\begin{enumerate}
\item
In Eq.(A.2), there appears a correction term to the "standard oscillation 
period", $4\pi E_{(kl)}/|m_k^2-m_l^2|$, which appears in Eq.(2.10) or 
Eq.(2.13). Although this correction term vanishes in either of the 
equal-energy, equal-momentum and equal-velocity cases, it is rather 
artificial, in the framework of the wave-packet treatment presented in 
this paper, to assume $E_k=E_l$ or $\vec{v}_k=\vec{v}_l$.
\item
In Eq.(A.4), each of oscillating factors is multiplied by two 
suppression factors: one which corresponds to that already present 
in Eq.(A.2), and in Eq.(2.10) or Eq.(2.13) as well, and the other 
which gives rise to another necessary condition for neutrino oscllation 
to be significant: 
$|E_{[kl]}|/\sqrt{2(\vec{v}_k^2+\vec{v}_l^2)} \lesssim \sigma_{\nu}$.
\end{enumerate}
\setcounter{equation}{0}
\def\theequation{B.\arabic{equation}}
\subsection*{Appendix B. Some algebra relevant to Sec.4}
If Equation (4.1) is substituted, $\bra\vec{v}\ket_{k,l}$ and $\bra\vec{X}\ket_{k,l}$ (defined by Eq.(3.11)) and $\bra\vec{u},\vec{w}\ket_{k,l}$ (given by Eq.(3.10)) may be expressed as
% (B.1)
\begin{eqnarray}
&& \bra\vec{v}\ket_{k,l} ~=~ \bra\vec{v}\ket_{(kl)} \pm \frac{\sigma_{\nu}^2}{2\sigma^2}\vec{v}_{[kl]}, \qquad 
\bra\vec{X}\ket_{k,l} ~=~ \bra\vec{X}\ket_{(kl)} \mp \frac{\sigma_{\nu}^2}{2\sigma^2}\vec{v}_{[kl]}t_{\nu}, \no \\
&& \bra\vec{u},\vec{w}\ket_{k,l} ~=~ \bra\vec{u},~\vec{w}\ket_{(kl)} \pm (\vec{u},\vec{w})_{kl}' + \bra\vec{u},\vec{w}\ket_{kl}'',
\end{eqnarray}
where
% (B.2)
\begin{eqnarray}
\bra\vec{u},\vec{w}\ket_{(kl)} 
&=& \frac{1}{\sigma^4}\{\sigma_{\nu}^2 \sigma_{\mu}^2 \vec{u}_{[(kl)\mu]}\vec{w}_{[(kl)\mu]} + \sigma_{\nu}^2 \sigma_{\pi}^2 \vec{u}_{[(kl)\pi]}\vec{w}_{[(kl)\pi]} + \sigma_{\mu}^2 \sigma_{\pi}^2 \vec{u}_{[\mu\pi]}\vec{w}_{[\mu\pi]}\}, \no \\
\bra\vec{u},\vec{w}\ket_{kl}' &=& \frac{1}{2\sigma^4}\sigma_{\nu}^2\{\sigma_{\mu}^2 (\vec{u}_{[kl]}\vec{w}_{[(kl)\mu]}+\vec{u}_{[(kl)\mu]}\vec{w}_{[kl]}) + \sigma_{\pi}^2(\vec{u}_{[kl]}\vec{w}_{[(kl)\pi]}+\vec{u}_{[(kl)\pi]}\vec{w}_{[kl]})\}, \nonumber \\
\bra\vec{u},\vec{w}\ket_{kl}'' &=& \frac{1}{4\sigma^4}\sigma_{\nu}^2(\sigma_{\mu}^2+\sigma_{\pi}^2)\vec{u}_{[kl]}\vec{w}_{[kl]}.
\end{eqnarray}
Accordingly, $a_{k,l}$, $b_{k,l}$ and $c_{k,l}$ may be expressed as
% (B.3)
\begin{eqnarray}
a_{k,l} &=& a_{(kl)} \pm a_{kl}' + a_{kl}'', \no \\
b_{k,l} &=& b_{(kl)} \pm b_{kl}' + b_{kl}'', \no \\
c_{k,l} &=& c_{(kl)} \pm c_{kl}' + c_{kl}'', 
\end{eqnarray}
where
% (B.4)
\begin{eqnarray}
a_{(kl)} &=& \bra\vec{v},\vec{v}\ket_{(kl)}, \quad a_{kl}' ~=~ \bra\vec{v},\vec{v}\ket_{kl}', \quad a_{kl}'' ~=~ \bra\vec{v},\vec{v}\ket_{kl}'', \no \\
b_{(kl)} &=& \bra\vec{v},\vec{X}\ket_{(kl)}, \quad b_{kl}' ~=~ \bra\vec{v},\vec{X}\ket_{kl}', \quad b_{kl}'' ~=~ \bra\vec{v},\vec{X}\ket_{kl}'', \no \\
c_{(kl)} &=& \bra\vec{X},\vec{X}\ket_{(kl)}, \quad c_{kl}' ~=~ \bra\vec{X},\vec{X}\ket_{kl}', \quad c_{kl}'' ~=~ \bra\vec{X},\vec{X}\ket_{kl}''.
\end{eqnarray}
Substituting Eq.(B.3) into Eq.(4.2), which defines $\zeta_{kl}$, one may derive 
Eq.(4.6). On the other hand, substituting Eqs.(B.2) and (B.4) into 
$c_{(kl)}-a_{(kl)}t_{(kl)}^2$, one may derive
\begin{eqnarray*}
c_{(kl)}-a_{(kl)}t_{(kl)}^2 &=& \frac{1}{8\sigma^8a_{(kl)}}\sum_{\kappa,\lambda,\kappa',\lambda'=(kl),\mu,\pi}~\sum_{i,j=x,y,z} \no \\
&& \sigma_{\kappa}^2\sigma_{\lambda}^2\sigma_{\kappa'}^2\sigma_{\lambda'}^2(v_{[\kappa\lambda]i}X_{[\kappa'\lambda']j}-v_{[\kappa'\lambda']j}X_{[\kappa\lambda]i})^2.
\end{eqnarray*}
Equation (4.5) therefore implies Eq.(4.7). From this, it follows 
that\footnote
{
Note also that the following relations hold independently of Eq.(4.7):
% (B.5)
\begin{eqnarray}
b_{kl}''=-a_{kl}''t_{\nu}, \qquad c_{kl}''=a_{kl}''t_{\nu}^2. 
\end{eqnarray}
}
%
% (B.6)
\begin{eqnarray}
b_{(kl)} &=& -a_{(kl)}\overline{t}_0, \qquad \qquad c_{(kl)} ~=~ a_{(kl)}\overline{t}_0^2, \no \\
b_{kl}' &=& -\frac{1}{2}a_{kl}'(t_{\nu}+\overline{t}_0), \qquad c_{kl}' ~=~ a_{kl}'t_{\nu}\overline{t}_0, \no \\
t_{k,l} &=& \frac{1}{2}(t_{\nu}+\overline{t}_0) - \frac{a_{(kl)}-a_{kl}''}{2(a_{(kl)} \pm a_{kl}'+a_{kl}'')}(t_{\nu}-\overline{t}_0).
\end{eqnarray}
With the aid of Eqs.(B.5) and (B.6), one may verify that $\zeta_{kl}$, 
Eq.(4.6), reduces to $\overline{\zeta}_{kl}(t_{\nu}-\overline{t}_0)^2$ 
with $\overline{\zeta}_{kl}$ given by Eq.(4.8).\footnote
{
It is not difficult to verify that $\overline{\zeta}_{kl} > 0$.
}

Similarly, if Equation (4.1) and 
% (A.7)
\begin{eqnarray}
E_{k,l} &=& E_{(kl)} \pm \frac{1}{2}E_{[kl]}
\end{eqnarray}
are substituted, $\Delta E_{k,l}$ and $\Delta\tilde{E}_{k,l}$ (defined 
respectively by Eq.(3.9) and Eq.(3.14)) may be expressed as
% (B.8)
\begin{eqnarray}
\Delta E_{k,l} &=& \Delta E_{(kl)} \pm \frac{1}{2}E_{[kl]}, ~~\qquad~~ \Delta\tilde{E}_{k,l} ~=~ \Delta\tilde{E}_{(kl)} \pm \frac{1}{2}\Delta\tilde{E}_{kl}',
\end{eqnarray}
where
% (B.9)
\begin{eqnarray}
\Delta\tilde{E}_{kl}' ~=~ E_{[kl]} - \frac{\sigma_{\nu}^2}{\sigma^2}\vec{v}_{[kl]}\Delta\vec{p}.
\end{eqnarray}
Substituting Eqs.(B.3) and (B.8) into Eq.(4.12), which defines 
$\eta_{kl}$, one readily verifies Eq.(4.16). Substituting Eqs.(3.9), 
(B.1) and (B.8) into Eq.(3.20), one may derive Eq.(4.20), which, 
if Equations (B.5) and (B.6) are further substituted, gives Eq.(4.21).
%
%%%References%%%
%

%
\end{document}